\documentclass[longauth]{aa}

\usepackage{amsmath}
\usepackage[english]{babel}
\usepackage{graphicx}
\usepackage[utf8]{inputenc}
\usepackage{float}

\usepackage[pdfencoding=auto,psdextra]{hyperref}
\hypersetup{%
  colorlinks=true,%
  linkcolor=blue,%
  filecolor=magenta,%
  urlcolor=blue,%
  citecolor=blue%
}
\usepackage{xcolor}

\usepackage{graphicx}
\usepackage{txfonts}
\usepackage{lipsum}
\usepackage{lscape}
\usepackage{placeins}
\usepackage{pdflscape}
\usepackage{booktabs}
\usepackage{rotating}
\usepackage{natbib}
\defcitealias{Zou:2024}{Z24}

\usepackage[nolist]{acronym}

\usepackage[nameinlink,capitalise]{cleveref}
\crefname{section} {Sect.}   {Sects.}
\Crefname{section} {Section} {Sections}
\crefname{figure}  {Fig.}    {Figs.}
\Crefname{figure}  {Figure}  {Figures}
\crefname{equation}{Eq.}     {Eqs.}
\Crefname{equation}{Equation}{Equations}
\crefname{table}   {Table}   {Tables}
\crefname{appendix}{Appendix}{Appendices}

\usepackage{siunitx}

\newcommand{\typei}{Type 1\xspace}
\newcommand{\typeii}{Type 2\xspace}
\newcommand{\lx}{L_\mathrm{X}}
\newcommand{\luv}{L_{2500\,\angstrom}}
\newcommand{\unithnaught}{\mathrm{km}\,\mathrm{s}^{-1}\,\mathrm{Mpc}^{-1}}
\newcommand{\unitfx}{\ensuremath{\mathrm{erg}\,\mathrm{cm}^{-1}\,\mathrm{s}^{-1}}}
\newcommand{\unitlx}{\ensuremath{\mathrm{erg}\,\mathrm{s}^{-1}}}
\newcommand{\unitlnu}{\ensuremath{\mathrm{erg}\,\mathrm{s}^{-1}\,\mathrm{Hz}^{-1}}}
\newcommand{\micron}{\ensuremath{$\textmu$\mathrm{m}}}

\newcommand{\angstrom}{\ensuremath{\text{\AA}}}

\newcommand{\nvisitdri}{1441}

\newcommand{\ntruthagnctn}{164\,005}
\newcommand{\ntruthagnctk}{136\,949}
\newcommand{\ntruthgalaxy}{7\,147\,592}
\newcommand{\ntruthstar}{624\,637}
\newcommand{\ntruth}{8\,073\,183}

\newcommand{\logten}{\ensuremath{\log_{10}}}
\newcommand{\logquantity}[2]{\ensuremath{\logten {( #1 \,/\, #2 )}}\xspace}
\newcommand{\loglambdaSAR}  {\logquantity{\lambda_\mathrm{SAR}}{\mathrm{erg}\,\mathrm{s}^{-1}\,M_\odot^{-1}}}
\newcommand{\logMstar}      {\logquantity{M_\mathrm{star}}{M_\odot}}
\newcommand{\logMBH}        {\logquantity{M_\mathrm{BH}}{M_\odot}}

\newcommand{\logLopt}       {\logquantity{L_{5100\,\angstrom}}{\mathrm{erg}\,\mathrm{s}^{-1}}}
\newcommand{\logLX}         {\logquantity{L_\mathrm{X}}{\mathrm{erg}\,\mathrm{s}^{-1}}}
\newcommand{\logLbol}       {\logquantity{L_\mathrm{bol}}{\mathrm{erg}\,\mathrm{s}^{-1}}}
\newcommand{\EBV}           {\ensuremath{E_{B-V}}\xspace}

\newcommand{\plambdafull}    {\ensuremath{p(\lambda_\mathrm{SAR} \,|\, M_\mathrm{star}, z, T)}\xspace}
\newcommand{\plambdactnfull} {\ensuremath{p_\mathrm{CTN}(\lambda_\mathrm{SAR} \,|\, M_\mathrm{star}, z, T)}\xspace}
\newcommand{\plambdactkfull} {\ensuremath{p_\mathrm{CTK}(\lambda_\mathrm{SAR} \,|\, M_\mathrm{star}, z, T)}\xspace}
\newcommand{\SNR}            {\ensuremath{\mathrm{S/N}}\xspace}

\usepackage{xspace}
\newcommand*{\Euclid}{\textit{Euclid}\xspace}

\newcommand*{\Chandra}{\textit{Chandra}\xspace}
\newcommand*{\eROSITA}{\textit{eROSITA}\xspace}
\newcommand*\XMMN{XMM-\textit{Newton}\xspace}
\newcommand*{\WMAP}{\textit{Wilkinson} Microwave Anisotropy Probe\xspace}

\newcommand{\sfont}[1]{{\scriptscriptstyle\rm #1}}
\newcommand{\IE}{\ensuremath{I_\sfont{E}}\xspace}
\newcommand{\YE}{\ensuremath{Y_\sfont{E}}\xspace}
\newcommand{\JE}{\ensuremath{J_\sfont{E}}\xspace}
\newcommand{\HE}{\ensuremath{H_\sfont{E}}\xspace}

\newcommand*{\arcsecf}{\hbox{$.\!\!^{\prime\prime}$}}
\newcommand*{\arcminf}{\hbox{$.\!\!^{\prime}$}}
\newcommand*{\degf}{\hbox{$.\!\!^{\circ}$}}

\newcommand{\eg}{e.g.,\xspace}
\newcommand{\ie}{i.e.,\xspace}

\begin{acronym}
  \acro{ACF}{autocorrelation function}
  \acro{AGILE}{AGNs In the LSST Era}
  \acro{AGN}{active galactic nucleus}
  \acroplural{AGN}[AGNs]{active galactic nuclei}
  \acro{BHMF}{black hole mass function}
  \acro{CC}{Classical Cepheid}
  \acro{CTN}{Compton-thin}
  \acro{CTK}{Compton-thick}
  \acro{DESC}{Rubin-LSST Dark Energy Science Collaboration}
  \acro{DIA}{difference image analysis}
  \acro{DDF}{Deep-Drilling Field}
  \acro{DRW}{damped random walk}
  \acro{DR1}{data release 1}
  \acro{DP0.2}{Data Preview 0.2}
  \acro{DP1}{Data Preview 1}
  \acro{EGG}{Empirical Galaxy Generator}
  \acro{ECDFS}{Extended \Chandra Deep Field South}
  \acro{FOV}{field-of-view}
  \acro{IGM}{intergalactic medium}
  \acro{INAF}{Istituto Nazionale di Astrofisica}
  \acro{LSST}{Vera C. Rubin Observatory Legacy Survey of Space and Time}
  \acro{LPV}{long-period variable}
  \acro{LRD}{little red dot}
  \acro{MAF}{Metrics Analysis Framework}
  \acro{ML}{machine learning}
  \acro{NMAD}{normalized median absolute deviation}
  \acro{MAD}{median absolute deviation}
  \acro{NIR}{near infrared}
  \acro{NUV}{near ultraviolet}
  \acro{MIR}{mid-infrared}
  \acro{OpSim}{Operations Simulator}
  \acro{RMS}{root-mean-square}
  \acro{SED}{spectral energy distribution}
  \acro{SF}{structure function}
  \acro{SFR}{star-formation rate}
  \acro{SMBH}{supermassive black hole}
  \acro{SMF}{stellar mass function}
  \acro{PSF}{point-spread function}
  \acro{QSO}{quasi-stellar object}
  \acro{UV}{ultraviolet}
  \acro{WFD}{Wide Fast Deep}
  \acro{XLF}{X-ray luminosity function}
\end{acronym}


\begin{document}

\title{AGILE: an end-to-end Rubin-LSST simulation of AGNs, galaxies, and stars}
\subtitle{I. Software description and first data release}

\author{%
  A.\ Viitanen\inst{\ref{inst:inaf_oar},\ref{inst:unige},\ref{inst:uh}}
  \and A.\ Bongiorno\inst{\ref{inst:inaf_oar}}
  \and I.\ Saccheo\inst{\ref{inst:bristol},\ref{inst:inaf_oar}}
  \and A.\ Grazian\inst{\ref{inst:inaf_oapd}}
  \and M.\ Paolillo\inst{\ref{inst:unina},\ref{inst:inaf_oacn}}
  \and V.\ Petrecca\inst{\ref{inst:inaf_oacn}}
  \and D.\ De Cicco\inst{\ref{inst:unina},\ref{inst:inaf_oacn}}
  \and D.\ Roberts\inst{\ref{inst:southampton}}
  \and F.\ Shankar\inst{\ref{inst:southampton}}
  \and V.\ Allevato\inst{\ref{inst:inaf_oacn},\ref{inst:sns}}
  \and E.\ Merlin\inst{\ref{inst:inaf_oar}}
  \and D.\ Ili\'c\inst{\ref{inst:belgrade},\ref{inst:hamburg}} %orcid=0000-0002-1134-4015
  \and A.\ B.\ Kova\v cevi\'c\inst{\ref{inst:belgrade}} %orcid=0000-0001-5139-1978
  \and G.\ De Somma\inst{\ref{inst:inaf_oacn},\ref{inst:infn_naples}}
  \and M.\ Di Criscienzo\inst{\ref{inst:inaf_oar}}
  \and L.\ Girardi\inst{\ref{inst:inaf_oapd}}
  \and M.\ Marconi\inst{\ref{inst:inaf_oacn}}
  \and A.\ Mazzi\inst{\ref{inst:unibo},\ref{inst:inaf_oapd}}
  \and G.\ Pastorelli\inst{\ref{inst:unipadova},\ref{inst:inaf_oapd}}
  \and M.\ Trabucchi\inst{\ref{inst:unipadova},\ref{inst:inaf_oapd}}
  \and T.\ Ananna\inst{\ref{inst:waynestate}}
  \and R.\ J.\ Assef\inst{\ref{inst:diegoportales}}
  \and W.\ N.\ Brandt\inst{\ref{inst:pennstate_astro},\ref{inst:pennstate_grav},\ref{inst:pennstate_phy}} % 0000-0002-0167-2453
  \and M.\ Brescia\inst{\ref{inst:unina}}
  \and A.\ W.\ Graham\inst{\ref{inst:swinburne}}
  \and G.\ Li\inst{\ref{inst:kavli_peking},\ref{inst:nao_china},\ref{inst:diegoportales}}
  \and D.\ Marsango\inst{\ref{inst:ufsm}}
  \and A.\ Peca\inst{\ref{inst:eureka},\ref{inst:yale}}
  \and M.\ Polioudakis\inst{\ref{inst:unina}}
  \and C.\ M.\ Raiteri\inst{\ref{inst:inaf_oato}}
  \and B.\ Rani\inst{\ref{inst:goddard},\ref{inst:maryland},\ref{inst:korea_space}}
  \and C.\ Ricci\inst{\ref{inst:unige},\ref{inst:diegoportales}}
  \and G.\ Richards\inst{\ref{inst:drexel}}
  \and M.\ Salvato\inst{\ref{inst:mpe},\ref{inst:origins}}
  \and S.\ Satheesh-Sheeba\inst{\ref{inst:diegoportales}}
  \and R.\ Shirley\inst{\ref{inst:mpe}}
  \and S.\ Tang\inst{\ref{inst:southampton}}
  \and M.\ J.\ Temple\inst{\ref{inst:durham}}
  \and F.\ Tombesi\inst{\ref{inst:roma2},\ref{inst:inaf_oar},\ref{inst:infn_roma}}
  \and I.\ Yoon\inst{\ref{inst:nrao},\ref{inst:virginia}}
  \and F.\ Zou\inst{\ref{inst:michigan}}
}

\institute{%
  INAF--Osservatorio Astronomico di Roma, Via Frascati 33, 00078 Monteporzio Catone, Italy \email{akke.viitanen@iki.fi}\label{inst:inaf_oar}
  \and Department of Astronomy, University of Geneva, ch.\ d’Ecogia 16, 1290 Versoix, Switzerland\label{inst:unige}
  \and Department of Physics, Gustaf Hällströmin katu 2, 00014 University of Helsinki, Finland\label{inst:uh}
  \and School of Physics, HH Wills Physics Laboratory, University of Bristol, Tyndall Avenue, Bristol, BS8 1TL, UK\label{inst:bristol}
  \and INAF--Osservatorio Astronomico di Padova, Vicolo dell'Osservatorio 5, I-35122, Padova, Italy\label{inst:inaf_oapd}
  \and Dipartimento di Fisica ``Ettore Pancini'' Università di Napoli Federico II, Via Cintia 80126, Italy\label{inst:unina}
  \and INAF--Osservatorio Astronomico di Capodimonte, Via Moiariello 16, 80131 Napoli, Italy\label{inst:inaf_oacn}
  \and School of Physics and Astronomy, University of Southampton, Highfield, Southampton SO17 1BJ, UK\label{inst:southampton}
  \and Scuola Normale Superiore, Piazza dei Cavalieri 7, 56126 Pisa, Italy\label{inst:sns}
  \and University of Belgrade-Faculty of Mathematics, Department of Astronomy, Studentski trg 16, 11000 Belgrade, Serbia\label{inst:belgrade}
  \and Hamburger Sternwarte, Universit\"at Hamburg, Gojenbergsweg 112, D-21029 Hamburg, Germany\label{inst:hamburg}
  \and Istituto Nazionale di Fisica Nucleare (INFN) -- Sez.\ di Napoli, Compl.\ Univ.\ di Monte S.\ Angelo, Edificio G, Via Cinthia, I-80126 Napoli, Italy\label{inst:infn_naples}
  \and Dipartimento di Fisica e Astronomia ``Galileo Galilei'', Università di Padova, Vicolo dell’Osservatorio 3, I-35122 Padova, Italy\label{inst:unipadova}
  \and Department of Physics and Astronomy ``Augusto Righi'', University of Bologna, via Gobetti 93/2, 40129 Bologna, Italy\label{inst:unibo}
  % Ananna
  \and Department of Physics and Astronomy, Wayne State University, Detroit, MI 48202, USA\label{inst:waynestate}
  % Assef
  \and Instituto de Estudios Astrofísicos, Facultad de Ingeniería y Ciencias, Universidad Diego Portales, Av. Ejército Libertador 441, Santiago, Chile\label{inst:diegoportales}
  % Brandt
  \and Department of Astronomy and Astrophysics, 525 Davey Laboratory, The Pennsylvania State University, University Park, PA 16802, USA\label{inst:pennstate_astro}
  \and Institute for Gravitation and the Cosmos, The Pennsylvania State University, University Park, PA 16802, USA\label{inst:pennstate_grav}
  \and Department of Physics, 104 Davey Laboratory, The Pennsylvania State University, University Park, PA 16802, USA\label{inst:pennstate_phy}
  % Graham
  \and Centre for Astrophysics and Supercomputing, Swinburne University of Technology, Hawthorn, VIC 3122, Australia\label{inst:swinburne}
  % Li
  \and Kavli Institute for Astronomy and Astrophysics, Peking University, Beijing 100871, People’s Republic of China\label{inst:kavli_peking}
  \and National Astronomical Observatories, Chinese Academy of Sciences, 20A Datun Road, Beijing 100101, People’s Republic of China\label{inst:nao_china}
  % Marsango
  \and Universidade Federal de Santa Maria (UFSM), Centro de Ci\^encias Naturais e Exatas (CCNE), Santa Maria, 97105-900, RS, Brazil\label{inst:ufsm}
  % Peca
  \and Eureka Scientific, 2452 Delmer Street, Suite 100, Oakland, CA 94602-3017, USA\label{inst:eureka}
  \and Department of Physics, Yale University, P.O. Box 208120, New Haven, CT 06520, USA\label{inst:yale}
  % Raiteri
  \and INAF, Osservatorio Astrofisico di Torino, Via Osservatorio 20, I-10025 Pino Torinese, Italy\label{inst:inaf_oato}
  % Rani
  \and NASA Goddard Space Flight Center, Greenbelt, MD 20771, USA\label{inst:goddard}
  \and Center for Space Science and Technology, University of Maryland Baltimore County, MD 21250, USA\label{inst:maryland}
  \and Korea Astronomy and Space Science Institute, 776 Daedeokdae-ro, Yuseong-gu, Daejeon 30455, Republic of Korea\label{inst:korea_space}
  % Ricci
  % Richards
  \and Department of Physics, Drexel University, 32 S. 32nd Street, Philadelphia, PA, 19104 USA\label{inst:drexel}
  % Salvato
  \and Max-Planck-Institut für extraterrestrische Physik, Giessenbachstr. 1, 85748 Garching, Germany\label{inst:mpe}
  \and Exzellenzcluster ORIGINS, Boltzmannstr. 2, D-85748 Garching, Germany\label{inst:origins}
  % Tang
  % Temple
  \and Centre for Extragalactic Astronomy, Department of Physics, Durham University, South Road, Durham DH1 3LE, UK\label{inst:durham}
  % Tombesi
  \and Physics Department, Tor Vergata University of Rome, Via della Ricerca Scientifica 1, 00133 Rome, Italy\label{inst:roma2}
  \and INFN -- Rome Tor Vergata, Via della Ricerca Scientifica 1, 00133 Rome, Italy\label{inst:infn_roma}
  % Yoon
  \and National Radio Astronomy Observatory, 520 Edgemont Rd, Charlottesville, VA 22903, USA\label{inst:nrao}
  \and Department of Astronomy, University of Virginia, P.O. Box 3818, Charlottesville, VA 22903, USA\label{inst:virginia}
  % Zou
  \and Department of Astronomy, University of Michigan, 1085 South University, Ann Arbor, MI 48109, USA\label{inst:michigan}
}

\date{Received Month DD, YYYY}

\abstract%
{%
}
{%
  Contemporary large-scale surveys such as the \ac{LSST} and Euclid
  present an unprecedented discovery potential for studying \acp{AGN} at the
  population level in the big data era. However, one major challenge is the
  accurate identification and classification of \acp{AGN} from
  optical and near-infrared photometry, or variability data alone. In
  order to optimize \ac{AGN} selection, classification, and systematics, as
  well as to test different data analysis tools, we present \acs{AGILE}
  (\acl{AGILE}), an \ac{LSST} end-to-end simulation software. \acs{AGILE} --
  developed as part of the INAF \ac{LSST} in-kind contribution -- is capable of
  simulating the anticipated AGN population in \ac{LSST} and Euclid.
}
{%
  We based \acs{AGILE} on existing simulations of galaxies and stars, while we
  developed an \ac{AGN} recipe based on empirical relations. \acs{AGILE}
  populates complete galaxy samples with \acp{AGN} according to the observed
  \ac{AGN} accretion rate distribution, and each \ac{AGN} is assigned an
  optical/UV spectral energy distribution. Optical \ac{AGN} variability is
  added using a damped random walk model connected to the \ac{AGN} physical
  parameters. Finally, \acs{AGILE} creates both \ac{LSST}-like images and
  related data products.
}
{%
  Using \acs{AGILE}, we build a $24\,\mathrm{deg}^2$ complete mock truth catalog
  of \acp{AGN}, galaxies, and stars with $0.2 < z < 5.5$, $\logMstar > 8.5$
  (\acp{AGN} and galaxies), and $r < 27.5\,\mathrm{mag}$ (stars). We
  also perform a pilot simulation (\acs{AGILE} DR1) consisting of
  $1\,\mathrm{deg}^2$ of \ac{LSST} operations in the COSMOS field observed up
  to three years in accordance with the survey strategy. We use \acs{AGILE} DR1
  to quantify the accuracy of the LSST Science Pipelines in recovering the true
  fluxes of \acp{AGN}, galaxies, and stars. We quantify the \ac{LSST}
  completeness and purity in recovering \typei \acp{AGN} using typical
  color-color and variability selections. We share the \acs{AGILE} DR1 dataset,
  an ideal test-bench for further scientific exploitation and forecasts in the
  context of \ac{LSST} \acp{AGN}.
}
{%
}


\maketitle

\acresetall

\section{Introduction}

The presence of a \ac{SMBH} with a mass in the range $\logMBH \approx 5$--$10$
at the centers of massive galaxies is an accepted paradigm in astronomy
\citep[\eg][]{kormendy_richstone95,kormendy_bender11}. These black holes grow
by mergers and accretion of gas and dust, the latter of which gives rise to the
highly energetic phenomena observed in \aclp{AGN}
\citep[\acsp{AGN}\acused{AGN}; \eg][]{lynden_bell69,alexander_hickox12}. Given
the strong correlations between the properties of galaxies and their central
\ac{SMBH}s
(\citealp{magorrian98,ferrarese_merritt00,kormendy13,graham2016ASSL..418..263G},
but see also \citealp{maiolino2024A&A...691A.145M} for the $z>4$ Universe),
understanding the role of \acp{AGN} is crucial for building a comprehensive
picture of galaxy evolution.

In the coming years, large-scale surveys such as
\acl{LSST} \citep[\acs{LSST}\acused{LSST}; ][]{ivezic2019ApJ...873..111I},
\Euclid \citep{euclid_mellier2025A&A...697A...1E},
\eROSITA \citep{merloni2012arXiv1209.3114M},
and others will usher \ac{AGN} science into the era of big data. These surveys
are expected to detect \acp{AGN} in the tens of millions, vastly expanding on
existing samples. However, a major challenge is to identify these \acp{AGN}
among the billions of sources (primarily galaxies) detected by these surveys.
Reliable \ac{AGN} selection based on optical and near-infrared
(NIR\acused{NIR}) photometry and variability will be a fundamental goal
\citep[\eg][]{Savic:2023,euclid_matamoro2025arXiv250315320E}. To prepare for
these surveys, it is essential to develop synthetic datasets that can serve as
test-beds for selection methodologies, completeness estimates, and data
analysis tools such as variability detection, morphology classification, and
photometric redshift estimation.

\ac{LSST} will be a ten-year optical survey covering approximately
$18\,000\,\mathrm{deg}^2$ of the southern extragalactic sky. With six
photometric bands ($ugrizy$) and hundreds of repeated observations per region,
the high-cadence imaging of \ac{LSST} will be a transformative tool
for \ac{AGN} research, particularly for variability studies. While \ac{LSST}
is expected to select a high-purity sample of some ten
million \acp{AGN} in the optical regime, \acp{AGN} detected by
\ac{LSST} is at least an order of magnitude larger
\citep[][]{LSST-Science-Collaboration:2009,li2025arXiv251208654L}. At the same
time, \Euclid will provide a complementary view of the \ac{LSST} sky with
space-based optical/\ac{NIR} imaging ($\IE < 24.5$, $\HE < 24$) and \ac{NIR}
spectroscopy ($R = 380$). This is especially helpful in uncovering the obscured
\ac{AGN} population
\citep[\eg][]{%
  euclid_matamoro2025arXiv250315320E,%
  euclid_tarsitano2025arXiv250315319E%
}.

In preparation for the massive \ac{LSST} dataset, we introduce
\ac{AGILE}, an end-to-end simulation pipeline developed as part of the
Italian \ac{INAF} in-kind contribution to \ac{LSST}. \ac{AGILE} is designed to
generate realistic \ac{AGN} catalogs and simulate their observational
properties as seen by \ac{LSST}, enabling a robust testing and optimization of
\ac{AGN} selection and classification strategies, as well as of a broad range
of data analysis tools. A similar work was conducted by the
\acl{DESC}\acused{DESC}
\citep[DESC;][]{korytov2019ApJS..245...26K,lsst2021ApJS..253...31L}; however,
it does not account for \acp{AGN} or high-redshift galaxies. The \ac{AGILE}
software is designed to complement this work by providing a complete census of
the \ac{AGN} population.

In \cref{sec:overview}, we present an overview of \ac{AGILE}.
\Cref{sec:galaxy} describes the generation of the galaxy catalogs,
followed by the \ac{AGN} population model in \cref{sec:agn} and the stellar
catalog in \cref{sec:star}. We show the validation results of the \ac{AGN}
catalog in \cref{sec:validation}, and the selected \ac{AGN} and stellar
variability recipes in \cref{sec:variability}. \Cref{sec:image} describes the
\ac{LSST}-like image simulations and \cref{sec:pipeline} outlines the
generation of photometric catalogs. Science applications are discussed in
\cref{sec:science}, followed by the summary and conclusions in
\cref{sec:conclusions}. Specifically, \cref{app:dr1} presents the
\ac{AGILE} \ac{DR1}. The \ac{AGILE} \ac{DR1} consists of a $24\,\mathrm{deg}^2$
truth catalog, and a total of $\nvisitdri$ simulated visits in the $ugrizy$
bands for a total of $21$ (out of a total of $189$) LSSTCam detectors covering
the central $1\,\mathrm{deg}^2$ for the first three years of the survey.
Throughout this work, we assume a flat $\Lambda$CDM cosmology with
$\Omega_\mathrm{m} = 0.3$ and
$H_0=70\,\unithnaught$
\citep[\WMAP;][]{spergel2003ApJS..148..175S},\footnote{%
  The most recent Planck measurements from the Cosmic Microwave Background
  suggest $H_0 = (67.4 \pm 0.5)\,\unithnaught$ and $\Omega_\mathrm{m} = 0.315
  \pm 0.007$ \citep{planck2020A&A...641A...6P}. However, the \WMAP cosmology is
  still assumed by many contemporary galaxy evolution studies.%
}
and the \citet{chabrier2003PASP..115..763C} initial mass function. Magnitudes
are expressed in the AB system \citep{oke1983ApJ...266..713O}.

\section{AGILE overview}%
\label{sec:overview}

\ac{AGILE} first builds a mock catalog including \acp{AGN}, galaxies, and stars
based on empirical relations, ensuring consistency with observed \ac{AGN} and
galaxy properties. The strength of the empirical approach is to accurately
reproduce the known underlying \ac{AGN} population, which can then be
directly compared to the recovered \ac{AGN} population by the \ac{LSST}
strategy, primarily characterized by its large survey area and cadence.
Moreover, this method follows the well-established methodology in the
literature, which in recent years has been successful in explaining
the large-scale clustering of \acp{AGN} in the context of selection effects
from X-ray surveys
\citep{%
  comparat2019MNRAS.487.2005C,%
  aird2021MNRAS.502.5962A,%
  allevato2021ApJ...916...34A,%
  viitanen2021MNRAS.507.6148V,%
  Lopez-Lopez:2024%
}.
Here, instead we focus on the optical properties of X-ray \acp{AGN} and
quasars\footnote{%
  We make no strict distinction between the terms \ac{AGN} and quasar and
  consider quasars as the high-luminosity and optically unobscured
  sub-population of \acp{AGN}.
}
in the context of \ac{LSST}, simulating as accurately as possible the effects
of both the hardware (the telescope and the LSSTCam instrument) and the
software (image processing pipelines). Therefore, \ac{AGILE} includes a model
for the instrumental effects, survey design, and time-domain variability for
all sources. This enables a dynamic, evolving representation of the \ac{AGN}
population as it will be observed by \ac{LSST}, providing an essential
framework for optimizing \ac{AGN} detection and classification.

\Cref{fig:agile} shows the full \ac{AGILE} flowchart, which can be divided into
three main steps: the first step involves generating an empirically motivated
mock truth catalog, where galaxies host X-ray \acp{AGN} following the latest
accretion rate distribution by \citet[][hereafter Z24]{Zou:2024}.
Further optical and \ac{UV} properties are assigned using empirical recipes
\citep[\eg][]{Lusso:2010,Merloni:2014}, and full \ac{UV}, optical, and \ac{NIR}
\acp{SED} are generated with \textsc{QSOGEN} \citep{Temple:2021b}. These
\ac{AGN} \acp{SED} are then combined with the galaxy \acp{SED} generated by
\textsc{EGG} \citep[Empirical Galaxy Generator;
][]{schreiber2017A&A...602A..96S} to derive photometry in the
optical/\ac{NIR}, including the \ac{LSST} and \Euclid bands. The
second step involves creating time-dependent instance catalogs,
where \ac{AGN} and stellar variability are captured by the \ac{LSST}
observing cadence. Finally, the instance catalogs are processed through the
\ac{LSST} image simulations, generating realistic \ac{LSST}-like single-visit
images. These images are then coadded and analyzed using the \ac{LSST} Science
Pipelines, mimicking real survey operations and producing final photometric
catalogs.

\section{AGILE mock galaxy catalog}%
\label{sec:galaxy}

The starting point of \ac{AGILE} is a complete (in terms of stellar
mass $M_\mathrm{star}$ and redshift $z$) population of galaxies. The
galaxy sample is created using \textsc{EGG}
\citep{schreiber2017A&A...602A..96S}, which is designed to generate realistic
mock catalogs of galaxies with physical properties such as
$M_\mathrm{star}$, \acp{SFR}, and \acp{SED}. \textsc{EGG} galaxies are
calibrated to reproduce the observed number counts within the GOODS
\citep{barro2019ApJS..243...22B,guo2013ApJS..207...24G} and CANDELS fields
\citep{grogin2011ApJS..197...35G,koekemoer2011ApJS..197...36K}, as well as the
angular two-point correlation function as a function of ${M_\mathrm{star}}$ and
$z$ using the \citet[][]{soneira1978AJ.....83..845S} algorithm. Moreover,
\textsc{EGG} galaxies are classified as either quiescent or star-forming based
on the galaxy \ac{SMF}, which is an input parameter.

For the \ac{AGILE} simulation of galaxies, we base the input galaxy \ac{SMF}
on the COSMOS $2\,\mathrm{deg}^2$ field \citep{scoville07}. The latest
COSMOS2020 catalog \citep{weaver2022ApJS..258...11W} has a comparable depth
to the \ac{LSST} wide survey after ten years \citep[$r \sim
27.5$;][]{bianco2022ApJS..258....1B}. Here, we use the quiescent and
star-forming galaxy \acp{SMF}, measured for $0.2 < z < 5.5$ galaxies which
are classified as quiescent and star-forming based on \ac{NUV}, $r$, and $J$
colors \citep{weaver2023A&A...677A.184W}. In the interest of having a complete
galaxy sample as the starting point, we allow for galaxies to be simulated down
to $\logMstar = 8.5$. This constitutes an extrapolation at higher $z$, as
COSMOS2020 $70\%$ mass completeness limits at $z = 1.2$ and $5.5$ are
$\logMstar = 8.5$ and $9.5$, respectively
\citep[][Sect.~3.3]{weaver2023A&A...677A.184W}.

\section{AGILE mock AGN catalog}
\label{sec:agn}

Following the empirical approach, we assign a realistic population of \acp{AGN}
to the complete population of galaxies according to the observed distributions
of \ac{AGN} properties and scaling relations
\citep[\eg][]{%
  georgakakis2019MNRAS.487..275G,%
  aird2021MNRAS.502.5962A,%
  allevato2021ApJ...916...34A,%
  viitanen2021MNRAS.507.6148V,%
  Lopez-Lopez:2024%
}.
Starting from a specific accretion rate ($\lambda_\mathrm{SAR} \equiv \lx \,/\,
M_\mathrm{star}$, where $\lx$ is the intrinsic $2$--$10\,\mathrm{keV}$ \ac{AGN}
luminosity) distribution function \citepalias{Zou:2024}, we build a complete
X-ray \ac{AGN} population over a wide luminosity baseline $L_\mathrm{X} \gtrsim
10^{41}\,\mathrm{erg\,s^{-1}}$ and $0.2 < z < 5.5$. We then convert it into a
population of optical/\ac{NIR} \acp{AGN} across different optical types (\ie
\typei or \typeii), and include \ac{AGN} optical variability.

\subsection{Accretion rate distribution}
\label{sec:accretion_rate_distribution}

\begin{figure*}[htbp]
  \centering
  \includegraphics[width=0.95\linewidth]{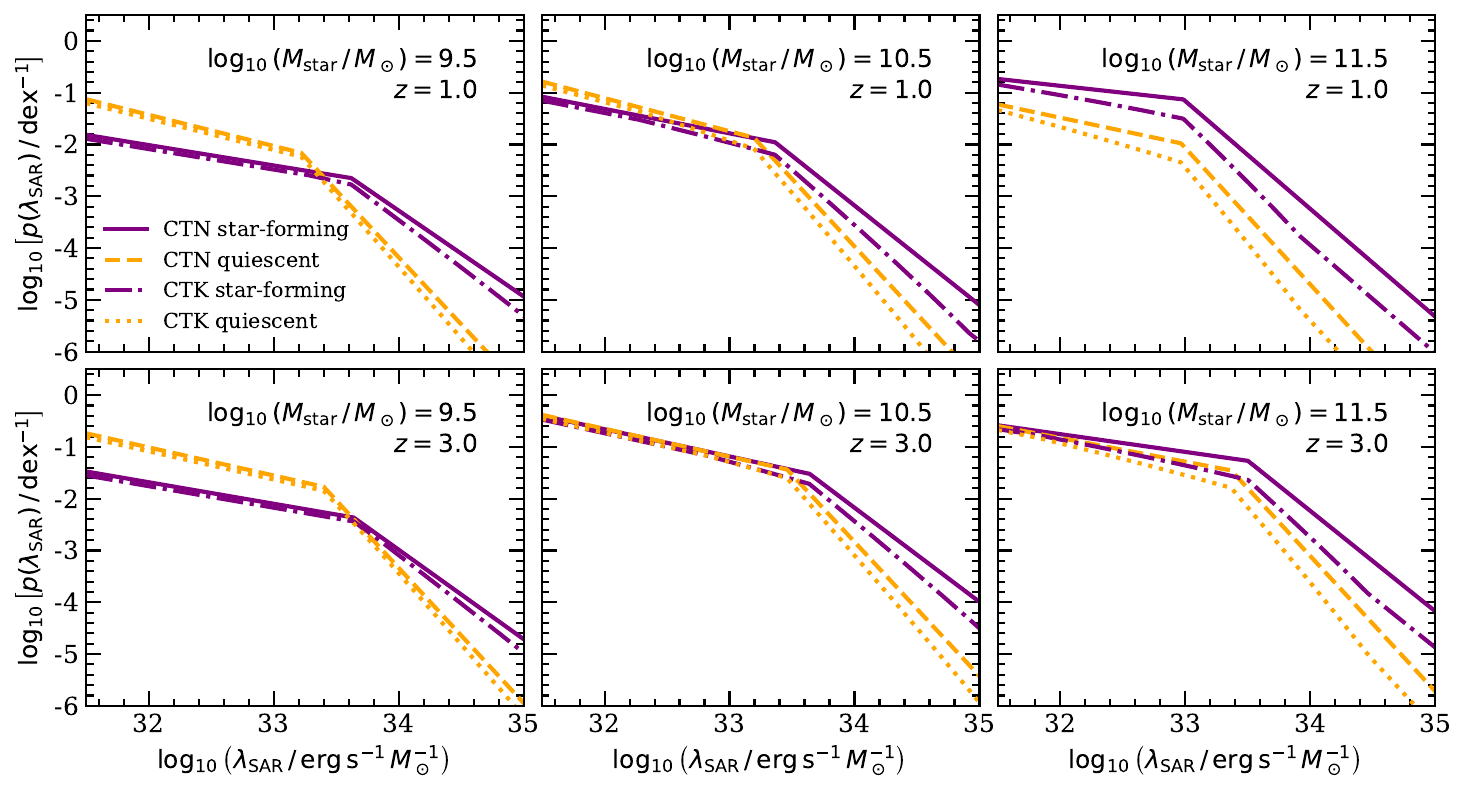}
  \caption{%
    Adapted distribution of $\lambda_\mathrm{SAR}$ of \acp{AGN}. Each panel
    shows $\plambdafull$ at different $M_\mathrm{star}$ (columns) and $z$
    (rows). The lines correspond to different combinations of \ac{AGN} host
    galaxy type (quiescent or star-forming), and \ac{AGN} obscuration (\acs{CTN}
    or \acs{CTK}) in accordance with the legend. The specific $M_\mathrm{star}$
    and $z$ shown here are selected for illustrative purposes, while the
    $\lambda_\mathrm{SAR}$ assignment follows the \citetalias{Zou:2024}
    parameter maps as explained in \cref{sec:accretion_rate_distribution}.
  }%
  \label{fig:pagn}
\end{figure*}

We start by assigning $\lambda_\mathrm{SAR}$ to each galaxy in the simulation
by using the \ac{AGN} $\lambda_\mathrm{SAR}$ distribution $\plambdafull$ (\ie
the probability that a quiescent or star-forming galaxy $T$ with
$M_\mathrm{star}$ at $z$ hosts an \ac{AGN} with $L_\mathrm{X}$), which is
observationally constrained in many studies
(\citealt{bongiorno2016A&A...588A..78B};
\citealt{georgakakis2017MNRAS.471.1976G};
\citealt{aird2018MNRAS.474.1225A};
\citealt{yang2018MNRAS.475.1887Y};
\citealt{laloux2024MNRAS.532.3459L};
\citetalias{Zou:2024}).
The $\lambda_\mathrm{SAR}$ distribution is strongly influenced by at least
${M_\mathrm{star}}$, $z$, and $T$, and there is evidence for further
dependencies on host-galaxy \ac{SFR}
\citep[\eg][]{aird2019MNRAS.484.4360A,yang2019MNRAS.485.3721Y}, compactness
\citep[\eg][]{ni2021MNRAS.500.4989N}, or on \ac{AGN} obscuration
\citep[\eg][]{ricci2017Natur.549..488R,laloux2024MNRAS.532.3459L}.

\subsubsection{Compton-thin AGNs}
\label{sec:ctnagn}

Here, we focus on reproducing the primary trends in the
\ac{AGN}-host galaxy connection in terms of $M_\mathrm{star}$, $z$, and
$T$, and use the most recent measurements of the \acl{CTN}
(\acs{CTN}\acused{CTN}, neutral hydrogen column density $N_\mathrm{H} <
10^{24}\,\mathrm{cm}^{-2}$) \plambdactnfull by \citetalias{Zou:2024}. Their
sample spans $10^{9.5} < {M_\mathrm{star}} / M_\odot < 10^{12}$ and $z < 4$,
and is derived from a large selection
of $8000$ \acp{AGN} and $1.3$ million
non-active galaxies compiled from nine different \XMMN, \Chandra, and \eROSITA
surveys with a wide reach
in the area-depth plane (from $0.05\,\mathrm{deg}^2\,/\,7000\,\mathrm{ks}$ to
$59.75\,\mathrm{deg}^2\,/\,2\,\mathrm{ks}$).
They find the observed \plambdactnfull well-fitted by a double power-law
with four free parameters: amplitude $A$, power-law slopes $\gamma_1$,
$\gamma_2$, and $\lambda_\mathrm{SAR,c}$ marking the transition
$\lambda_\mathrm{SAR}$ between the two power-law slopes.

For each galaxy in \ac{AGILE} with a given ${M_\mathrm{star}}$, $z$, and $T$,
we use the median model parameter maps shown in Fig.~2 of \citetalias{Zou:2024}
in order to find the corresponding parameters $A, \gamma_1, \gamma_2,
\lambda_\mathrm{SAR,c}$. Then, \plambdactnfull is given by
\begin{multline}
    \plambdactnfull \\
    =
    \begin{cases}
        A \times \left( \lambda_\mathrm{SAR} / \lambda_\mathrm{SAR,c} \right)^{-\gamma_1}, & \text{if } \lambda_\mathrm{SAR} <    \lambda_\mathrm{SAR,c} \\
        A \times \left( \lambda_\mathrm{SAR} / \lambda_\mathrm{SAR,c} \right)^{-\gamma_2}, & \text{if } \lambda_\mathrm{SAR} \geq \lambda_\mathrm{SAR,c}
    \end{cases},
\end{multline}
and we assign each galaxy a value of $\lambda_\mathrm{SAR}$, drawn at random
using the corresponding distribution. We extrapolate the parameter maps down to
$\logMstar = 8.5$ and up to $z = 5.5$ by taking the boundary values of the
parameter maps at $\logMstar = 9.5$ and $z = 4$, respectively. The choice of
the extrapolation scheme affects \plambdactnfull the most at $z \lesssim 0.5$
(up to $2.0\,\mathrm{dex}$), while the differences are $<0.5\,\mathrm{dex}$ at
$0.5 < z < 4$. We show these trends and discuss this further in
\cref{app:plambda_extrapolation}.

Recent observations suggest that the black-hole occupation fraction
$f_\mathrm{occ}$ \ie the probability of a galaxy hosting a \ac{SMBH} could
depend on $M_\mathrm{star}$
\citep{%
  miller2015ApJ...799...98M,%
  burke2025ApJ...978...77B,%
  zou2025ApJ...992..176Z%
}.
In particular, \citet{zou2025ApJ...992..176Z}, studying a sample of local
($<50\,\mathrm{Mpc}$) galaxies, measured  $f_\mathrm{occ}$ to be approximately
$30\%$, $60\%$, and $100\%$ at $\logMstar = 8.5$, $9.5$, and $10.5$,
respectively. Given that $f_\mathrm{occ}$ is observationally constrained only
in the local Universe, we do not account for it in the simulation where we
assume that every galaxy hosts a SMBH. However, $f_\mathrm{occ}$  is computed
and a flag is provided for each galaxy in the truth catalog (see
\cref{sec:validation}) and we verified that the validation results discussed in
\cref{sec:validation} hold also when $f_\mathrm{occ}$ is applied and
extrapolated to any redshift.

Following \citetalias{Zou:2024} and \citet{aird2018MNRAS.474.1225A}, in
\ac{AGILE} a galaxy is defined to host an \ac{AGN}, if $\lambda_\mathrm{SAR} >
10^{32} \,\mathrm{erg}\,\mathrm{s}^{-1}\,M_\odot^{-1}$. Assuming a bolometric
correction $L_\mathrm{bol}\,/\,L_\mathrm{X} = 25$, and
${M_\mathrm{star}}\,/\,{M_\mathrm{BH}} = 500$, this corresponds approximately
to an accretion rate of $1\%$ Eddington
\citep[][]{%
  aird2012ApJ...746...90A,%
  bongiorno2016A&A...588A..78B,%
  georgakakis2017MNRAS.471.1976G%
}.
Using this cut, the \ac{AGN} fraction is of the order of a few percent at $z <
1.0$ to few tens of percent at $z > 1.0$ \citepalias[][Fig.~7]{Zou:2024}, in
agreement with the observed \ac{AGN} fraction of
\citet{aird2018MNRAS.474.1225A}.

\subsubsection{Correcting for Compton-thick AGNs}

It is important to note that \citetalias{Zou:2024} derived results only for
\ac{CTN} \acp{AGN}. Indeed, while X-ray selection in the observed
$2$--$10\,\mathrm{keV}$ band is among the most efficient ways of selecting
\acp{AGN}, it is still biased against heavily obscured \ac{CTK} \acp{AGN} with
$N_\mathrm{H} > 10^{24} \,\mathrm{cm}^{-2}$. Given that X-ray background
population synthesis models suggest that $30$--$50\%$ of \acp{AGN} could be
\ac{CTK}
\citep{%
	ananna2019ApJ...871..240A,%
	peca2023ApJ...943..162P,%
	annuar2025MNRAS.540.3827A%
}
and mostly undetected by hard X-ray selection, we add in a correction for
\ac{CTK} \acp{AGN}.

In order to reproduce the \ac{CTK} \ac{AGN} number density as a function of
$L_\mathrm{X}$ and $z$, we use the \ac{AGN} obscuration distribution function.
Following \citet[][]{ueda2014ApJ...786..104U}, we define $f({N_\mathrm{H}}
\,|\, {L_\mathrm{X}}, z)$ (in units of $\mathrm{dex}^{-1}$) as the fraction of
\acp{AGN} at a given $N_\mathrm{H}$, $L_\mathrm{X}$, and $z$. Then, the
\ac{CTK} \ac{AGN} fraction is given by:
\begin{equation}
  \mathrm{frac}_\mathrm{CTK}({L_\mathrm{X}}, z)
  =
  \frac{\int_{24}^{26} f({N_\mathrm{H}} \,|\, {L_\mathrm{X}}, z) \, \mathrm{d} \logten N_\mathrm{H}}
       {\int_{20}^{26} f({N_\mathrm{H}} \,|\, {L_\mathrm{X}}, z) \, \mathrm{d} \logten N_\mathrm{H}}.
  \label{eq:fctk}
\end{equation}
Moreover, due to the lack of estimates of the accretion rate distribution of
\ac{CTK} \acp{AGN} ($p_\mathrm{CTK}$), given $\lambda_\mathrm{SAR}$,
${M_\mathrm{star}}$, $z$, and $T$, we assume that all \acp{AGN} are either
\ac{CTN} or \ac{CTK} ($p_\mathrm{AGN} = p_\mathrm{CTN} + p_\mathrm{CTK}$), and
that \ac{CTK} \acp{AGN} relate to the total \ac{AGN} population via
$p_\mathrm{CTK} = \mathrm{frac}_\mathrm{CTK}\,\times\,p_\mathrm{AGN}$.
Then, the \ac{CTK} \ac{AGN} accretion rate distribution is given by
\begin{multline}
  \plambdactkfull \\
  = \frac{\mathrm{frac}_\mathrm{CTK}({L_\mathrm{X}}, z)    }%
         {1 - \mathrm{frac}_\mathrm{CTK}({L_\mathrm{X}}, z)}
  \plambdactnfull
  ,
\end{multline}
where ${L_\mathrm{X}} = \lambda_\mathrm{SAR} \times {M_\mathrm{star}}$, and we
assume $p_\mathrm{CTN}$ from \citetalias{Zou:2024}. Consequently, the
combined \ac{AGN} accretion rate distribution is defined as
$p_\mathrm{AGN} = p_\mathrm{CTN} + p_\mathrm{CTK}$, and the total \ac{AGN}
fraction is given by the integral
\begin{multline}
    \mathrm{frac}_\mathrm{AGN}(\lambda_\mathrm{SAR} \,|\, {M_\mathrm{star}}, z, T) \\
    =
    \int_{\lambda_{\mathrm{SAR},\min}}
    p_\mathrm{AGN}(\lambda_\mathrm{SAR} \,|\, {M_\mathrm{star}}, z, T) \,
    \mathrm{d} \logten \lambda_\mathrm{SAR},
\end{multline}
where we define $\lambda_{\mathrm{SAR},\min} = 10^{32}
\,\mathrm{erg}\,\mathrm{s}^{-1}\,M_\odot^{-1}$. We show the resulting accretion
rate distributions of \ac{CTN} and \ac{CTK} \acp{AGN} in \cref{fig:pagn}. For
the \ac{CTK} \ac{AGN} occupation fraction (see \cref{sec:agn}), we choose to
account for it by employing an identical recipe as for the \ac{CTN}
\acp{AGN}\@.

\subsection{Supermassive black hole masses}
\label{sec:mbh}

\begin{figure*}[htbp]
  \centering
  \includegraphics[width=\linewidth]{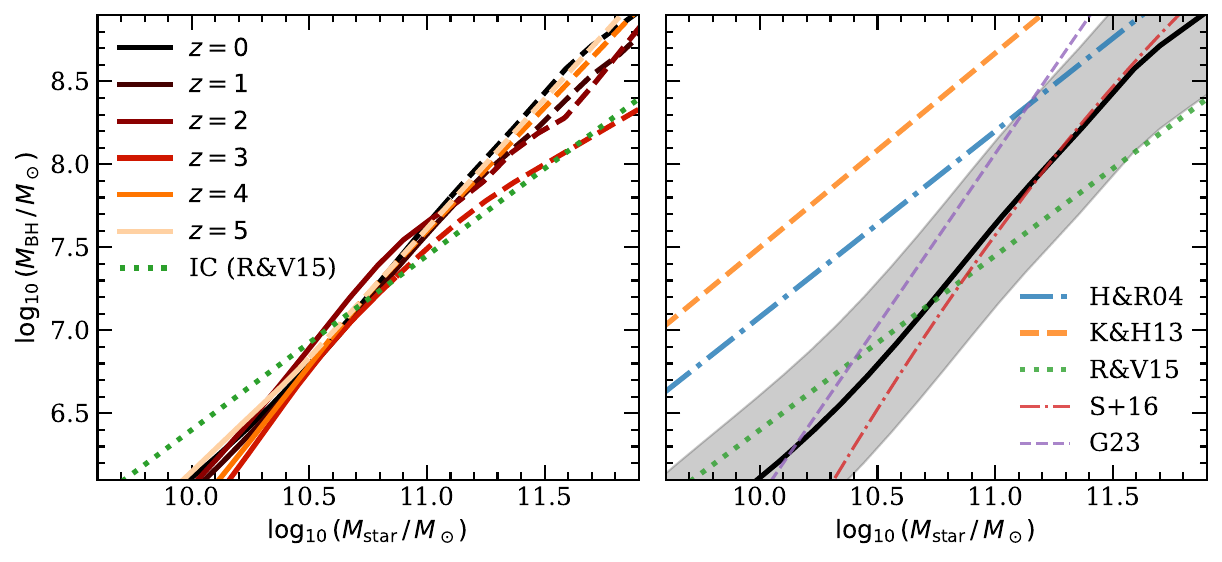}
  \caption{%
    The resulting $M_\mathrm{BH}$--$M_\mathrm{star}$ scaling relation from the
    continuity equation. The left panel shows the $z$ evolution of the scaling
    relations from $z=0$ (darker) to $z=5$ (lighter).
    At each redshift, the dashed line style indicates the regime above the
    $99\%$ stellar mass limits (assuming \citealt{weaver2023A&A...677A.184W}
    COSMOS2020 stellar mass function and an area of $24\,\mathrm{deg}^2$),
    above which the ${M_\mathrm{BH}}-{M_\mathrm{star}}$ relation is to be
    considered an extrapolation. The dotted line shows the assumed initial
    conditions at $z=5.5$ \citep{reines15}. The right panel shows the local
    relation implied by the continuity equation (black line) and the shaded
    region corresponds to an assumed scatter of $\Delta \logMBH =
    0.50\,\mathrm{dex}$. The other non-solid lines correspond to local and
    inactive early-type galaxies \citep{haring_rix04,kormendy13}, local
    \acp{AGN} \citep{reines15}, the de-biased relation from SDSS galaxies
    \citep{shankar16}, and major-merger built S0 and E galaxies
    \citep{graham2023MNRAS.522.3588G}.
  }%
  \label{fig:mbhmstar}
\end{figure*}

The mass of the \ac{SMBH} (${M_\mathrm{BH}}$) is one of the most fundamental
\ac{AGN} properties. Tight correlations are observed between ${M_\mathrm{BH}}$
and galaxy properties
\citep[velocity dispersion, luminosity, and $M_\mathrm{star}$;][]{%
  magorrian98,%
  gebhardt00,%
  ferrarese02,%
  kormendy13%
},
and discrepant scaling relations are known between early-type (merger-driven
evolution) and late-type galaxies
\citep[\eg][Fig.~A4]{graham2023MNRAS.522.3588G}. In order to assign each
\ac{AGN} a ${M_\mathrm{BH}}$, we assume the $M_\mathrm{BH}$--$M_\mathrm{star}$
relation derived by using a novel method based on the continuity equation. In
the continuity equation
\citep[\eg][]{yang2018MNRAS.475.1887Y,shankar2020MNRAS.493.1500S}, galaxies are
initially seeded with $M_\mathrm{BH}$ at $z=5.5$ following the local
scaling relation \citep{reines15}. Then, the growths of galaxies and BHs are
governed by the evolution of the \ac{SMF}
\citep[COSMOS2020;][]{weaver2023A&A...677A.184W} and $p(\lambda_\mathrm{SAR})$
\citepalias{Zou:2024}, respectively.
We assume that accretion is the dominant channel of BH growth and ignore BH
mergers which have a negligible impact apart from the local Universe and
high-mass BHs. We further assume a radiative and kinetic efficiency of $0.10$
and $0.05$, respectively
\citep[\eg][]{soltan1982MNRAS.200..115S,shankar2020NatAs...4..282S}.
The strength of this method is that it produces a self-consistent BH population
across different $z$, and the ${M_\mathrm{BH}}$--${M_\mathrm{star}}$ relation
is a direct output of the simulation. This approach is favored to assuming an
empirical $z$-dependent ${M_\mathrm{BH}}$--${M_\mathrm{star}}$ relation that
has the downside of including any biases which are present in that particular
measurement or galaxy population.
We use the output ${M_\mathrm{BH}}$--${M_\mathrm{star}}$ relation from the
continuity equation to assign each galaxy an $M_\mathrm{BH}$ inclusive of an
intrinsic scatter of ${\sim} 0.50\,\mathrm{dex}$.
\citep[][]{reines15,shankar19}.\footnote{%
  Recent JWST observations of overmassive \acp{AGN} at $z>4$ could imply a $z$
  evolution of the scatter \citep[\eg][]{li2025ApJ...981...19L} which we do not
  take into account in the interest of focusing on the majority of the lower
  $z$ \ac{AGN} population detectable by \ac{LSST}.
}
We show the resulting ${M_\mathrm{BH}}$--${M_\mathrm{star}}$ relation in
\cref{fig:mbhmstar}.

As can be seen, the $M_\mathrm{BH}$--$M_\mathrm{star}$ relation evolves rapidly
from $z=5.5$ to $5.0$, indicating that the continuity equation implies a
different overall $M_\mathrm{BH}$--$M_\mathrm{star}$ relation to the initial
conditions from \citet[][]{reines15}. This rapid evolution at these $z$ is a
simulation artifact as in the continuity equation approach the initial
conditions are quickly washed away. Then, the output
$M_\mathrm{BH}$--$M_\mathrm{star}$ relation settles quickly to the one implied
by the continuity equation. Therefore, at these $z$ the $M_\mathrm{BH}$ values
are to be considered unreliable. We further highlight the $z$-dependent 99\%
mass completeness limit (assuming COSMOS2020 and an area of
$24\,\mathrm{deg}^2$) in \cref{fig:mbhmstar} as dashed lines. In this regime,
the $M_\mathrm{BH}$--$M_\mathrm{star}$ relation may be considered an
extrapolation.

\subsection{Optical-UV properties: UV luminosity and AGN type}

\begin{figure*}
    \centering
    \includegraphics[width=0.9\linewidth]{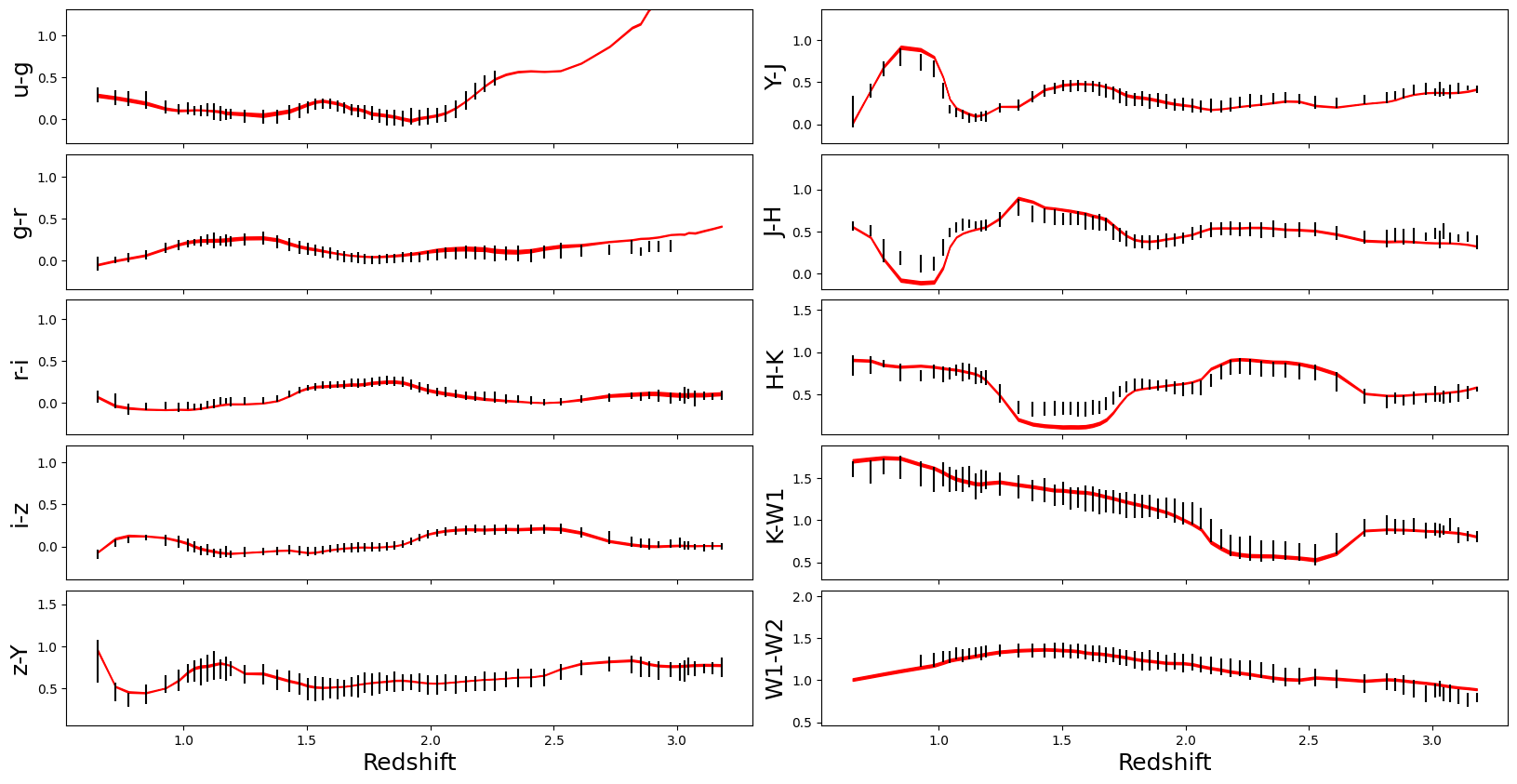}
    \caption{%
      Simulated (red) and observed SDSS DR16
      \citep[black;][]{ahumada2020ApJS..249....3A} colors versus $z$. Data and
      filters are from SDSS-DR16 \citep[$ugriz$,][]{Lyke:2020}, UKIDSS-LAS
      \citep[$YJHK$,][]{Lawrence:2012} and unWISE
      \citep[$W1W2$,][]{Schlafly:2019}. For each $z$ bin, $200$ combinations of
      parameters were drawn from the posterior, the thickness of the line
      denotes the $\pm 1\,\sigma$ region of the simulated colors.
    }
    \label{fig:mock_simulated_colors}
\end{figure*}

In order to derive realistic $ugrizy$ photometry for the \ac{AGN} population,
we describe here our optical/\ac{UV} model of the X-ray \ac{AGN} population.
First, we assign each \ac{AGN} a monochromatic \ac{UV} luminosity $\luv$
based on the well-established relationship between $\luv$ and
$L_{2\,\mathrm{keV}}$ \citep[\eg][]{Just:2007,Lusso:2010,Lusso:2016}.%
\footnote{%
  However, recent works \citep[\eg][]{Lopez-Lopez:2024} suggest discrepant
  terms for the local low-luminosity \acp{AGN}. Here, we focus on the $0.2 < z
  < 5.5$ Universe and our simulation aims to capture accurately the smaller
  area pencil-beam surveys. For the half-sky \ac{LSST}, local low-luminosity
  \acp{AGN} would have to be considered as a separate population.
}
Starting from the intrinsic $2$--$10\,\mathrm{keV}$ luminosity, we derive the
monochromatic $2\,\mathrm{keV}$ luminosity, assuming a standard power-law
shape with $\Gamma = 1.9$
\citep[\eg][]{cappelluti2009A&A...497..635C}.
Then, we use the
$\luv$--$L_{2\,\mathrm{keV}}$ relation
of \citet[][Eq.~5]{Lusso:2010} for \typei \acp{AGN} in $XMM$-COSMOS\@.
Finally, we factor in an intrinsic log-normal scatter of $0.40$ dex in the
$\luv$, consistent with \citet{Lusso:2010}.

Only a fraction of the X-ray \ac{AGN} population is expected to have
optical/\ac{UV} counterparts due to extinction by dust obscuration. To account
for this, we assign each \ac{AGN} an optical classification between \typei
(optically unobscured) and \typeii (optically obscured). Here, we use the
\typeii \ac{AGN} fraction as a function of $L_\mathrm{X}$ and $z$
based on $XMM$-COSMOS \citep[][]{Merloni:2014}.
Moreover, we classify conservatively all \ac{CTK} \acp{AGN} as
\typeii{}.
While the optical classification could be based on $N_\mathrm{H}$
\citep[\eg][]{ueda2014ApJ...786..104U}, \citet[][]{Merloni:2014} do report on
two \ac{AGN} populations where the optical and X-ray classifications do not
agree. These populations would be neglected in a scheme only based on
$N_\mathrm{H}$.

\subsection{Multi-wavelength AGN Spectral Energy Distribution}
\label{sec:qsogen}

To generate the \ac{AGN} \ac{SED} templates, we use \textsc{QSOGEN}
\citep{Temple:2021b}, which models the \ac{SED} as the sum of various
components representing the different \ac{AGN} physical processes with a total
of nine free parameters.
These templates represent time-averaged \acp{SED} and do not account for
\ac{AGN} variability, which we discuss in detail in
\cref{sec:agnvariability}.
To briefly summarize the role of the parameters, \textsc{QSOGEN} models the
accretion disk emission as a double power law, characterized by the spectral
indices \texttt{pslp1} and \texttt{pslp2}, with a break at the wavelength
\texttt{plbrk1}.\footnote{%
  Accretion disk emission is modeled with three power laws but the third one
  has a fixed break at $1216\,\angstrom$, and a fixed slope $\texttt{pslp3} =
  \texttt{pslp1} - 1$ so it does not belong to the free parameters.
}
The \ac{NIR} emission, originating from hot dust in the inner regions of the
torus, is modeled as a black body with temperature $T_\mathrm{BB}$ and
normalization \texttt{BB\_norm}. Broad emission lines are incorporated from
composite spectra, with their intensity and equivalent width scaled via the
parameters \texttt{scal\_emline} and \texttt{beslope}, respectively. The
host-galaxy emission is controlled by the parameters \texttt{fragal} and
\texttt{gpblind}, where \texttt{fragal} defines the \ac{AGN}-to-host flux
fraction, and \texttt{gpblind} its dependence on \ac{AGN} luminosity.

The posterior distribution of the free parameters is obtained by fitting the
colors of \acp{QSO} as a function of $z$. For each set of parameters, a
\ac{SED} is generated, from which synthetic colors are computed across
different $z$, accounting also for \ac{IGM} absorption
\citep{Becker:2013}. These synthetic colors are then fitted against the
observed data.

Notably, the best-fit parameters in \citet{Temple:2021b} are calibrated against
a mixed population of quasars and \acp{AGN} with significant host-galaxy
contamination. Instead, our chosen methodology of populating galaxies with
\acp{AGN} requires a pure \ac{AGN} \ac{SED} to be added on top of the galaxy
one. Thus, we followed a similar approach to the one used by
\citet{Buchner:2024} for the construction of the Chimera QSO benchmark, and
restricted the original sample to luminous quasars with $L_{5100\,\angstrom}
\geq 10^{45.5}\,\mathrm{erg}\,\mathrm{s}^{-1}$. This sample may be considered
devoid of host-galaxy contamination \citep[\eg]{Shen:2011}, and allow us to
derive pure \ac{AGN} \ac{SED} templates. We note that a potential limitation in
our approach is that our sample is biased toward luminous, high-Eddington
accreting sources. As a result the derived SED may not be fully representative
of lower luminosity, low-accretion rate AGN, which are known to exhibit
different SED shapes due to their distinct accretion regimes
\citep[e.g.][]{Lopez:2024}

We start with the \citet{Temple:2021b} \ac{QSO} sample from the SDSS Quasar
DR16 catalog \citep{Lyke:2020} cross-matched with the UKIDSS Large
Area Survey catalog \citep{Lawrence:2012}, and the unWISE catalog
\citep{Schlafly:2019}. Out of $95\,684$ \acp{QSO} which are detected in all the
bands (SDSS $ugriz$, UKIDSS $YJHK$, and unWISE $W1$--$2$), we find a final
sample size of $N=15\,790$ \acp{QSO} with $\logLopt \geq 45.5$ and high-quality
photometry ($\Delta m < 0.10$). The $z$ distribution of the final sample spans
across $0.6 \leq z \leq 3.2$. This $z$ cut ensures sufficient source density
across the covered $z$ interval, which is not the case at low-$z$ due to the
luminosity threshold and at very high-$z$ due to the $\Delta m < 0.1$
requirement.

We then find the posterior distribution of the best-fit \textsc{QSOGEN}
parameters by fitting the observed distribution of \ac{QSO} colors as a
function of $z$. Following \citet{Temple:2021b}, we fit the average colors
computed in different $z$ bins, which are designed to contain at least $30$
\acp{QSO} and to be spaced, on average, by $\Delta z = 0.035$. For each bin, we
compute the mean using a sigma clipping method with a threshold of $2\,\sigma$
to exclude significantly dust reddened sources. We also assign a scatter for
each bin, which is three times the standard deviation in that bin, in order to
obtain a final broader posterior distribution.

In the fit, we limit our analysis to the combination of bins and colors whose
filters fall within the rest-frame range from $912\,\angstrom$ to $3\,\micron$.
We removed the host-galaxy contribution by first disabling the
parameters \texttt{fragal} and \texttt{gpblind}. Moreover, our sample does not
allow us to reliably constrain the anti-correlation in the line strength and
\ac{AGN} luminosity (the Baldwin effect, \citealt{Baldwin:1977}) because of the
limited range of luminosities probed ($\logLopt \geq 45.5$).
Therefore, we chose to fix the broad emission lines intensity and their
luminosity dependence governing parameters (\texttt{beslope}, and
\texttt{scal\_emline}) to the values of \citet{Temple:2021b}. We show the
distribution of observed and synthetic colors and the posterior distribution of
the best-fit parameters in \cref{fig:mock_simulated_colors,fig:posterior},
respectively.

As shown in \cref{fig:mock_simulated_colors}, the average \ac{QSO} colors are
well represented, although some discrepancies are present for the $J-H$ and
$H-K$ colors in the $z$ ranges $0.8$--$1.2$ and $1.2$--$1.6$, respectively,
which approximately correspond to the wavelength interval
$6000\,\angstrom$--$1\,\micron$. In this range, our predicted colors are bluer
than the observed ones, likely due to a small residual host-galaxy
contamination not accounted for in our modeling.

We finalized our \ac{SED} templates by adding narrow emission lines to \typeii
\acp{AGN}, following the models by \citet{Feltre:2016} and using the same
approach and grid of parameters as in \citet{Lopez-Lopez:2024}. The narrow line
templates are then normalized according to the \ac{AGN} $\lx - L_{[\text{O
III}]}$ relation from \citet{Lamastra:2009}. We applied extinction the
templates, based on the host-galaxy optical depth provided by \textsc{EGG}, and
using the reddening law by \citet{Calzetti:2000}.

\begin{figure}
    \centering
    \includegraphics[width=1.0\linewidth]{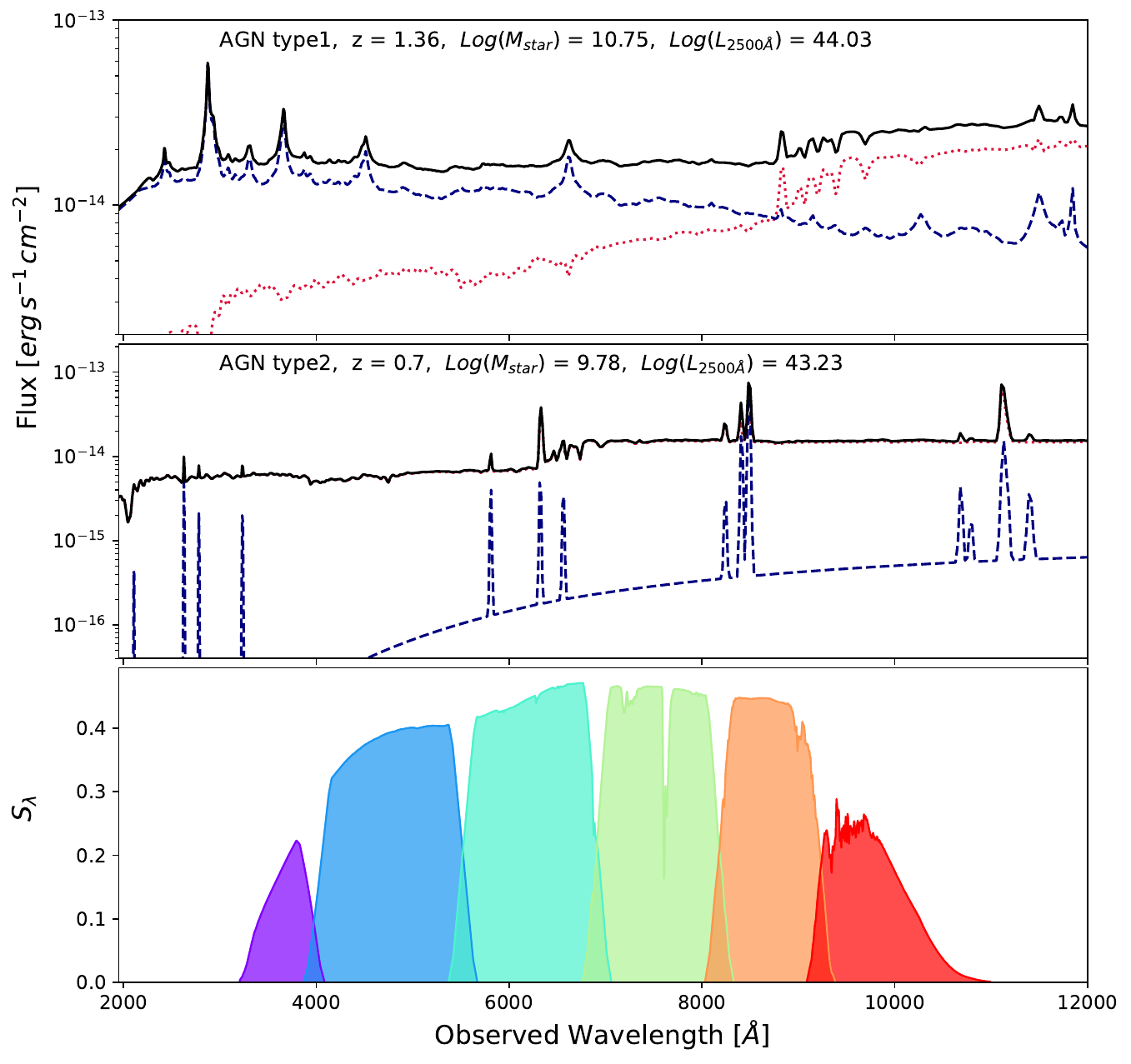}
    \caption{%
      Examples of \ac{AGN} (blue) and galaxy (red) \acp{SED} in the observer
      frame. The black line shows the combined \ac{SED}. The top (middle) panel
      shows a luminous \typei{} (\typeii{}) \ac{AGN}. The bottom panel shows
      the \ac{LSST} $ugrizy$ transmission curves.
    }
    \label{fig:example_SED}
\end{figure}

Using our updated \ac{SED} model, we then assign an optical/\ac{UV} \ac{SED} to
each \ac{AGN} by drawing a set of best-fit parameters from the posterior
distribution, and normalize the \ac{SED} to the value of $\luv$. We redden the
intrinsic \ac{SED} by assuming an optical/\ac{UV} extinction characterized by
the \ac{AGN} $\EBV$. We calibrate the distribution of $\EBV$ to that
suggested by \ac{SED} fitting of \typei{} X-ray \acp{AGN} in the \ac{LSST}
\aclp{DDF} \citep[\acsp{DDF}\acused{DDF}; ][]{Zou:2022}. Instead, for \typeii
\acp{AGN} whose extinction arises from a (partially) obscuring torus, we use
the observed $\EBV$ distribution from \typeii \acp{AGN} in $XMM$-COSMOS
\citep{bongiorno12}. In both cases, we find the observed distribution
well-represented by the functional form $p(\EBV) \propto {\left[ 1 + \beta^n \,
(\EBV)^n \right]}^{-1}$ \citep{hopkins:2004,Krawczyk:2015}, with $\beta =
15.20$ and $n = 1.58$. For \typeii \acp{AGN}, the data suggest an additional
offset of $+0.3$ in terms of $\EBV$. We apply $\EBV$ following the reddening
law described in \citet{Temple:2021b}. Examples of the final \acp{SED}
for \typei and \typeii \acp{AGN} -- including both the \ac{AGN} and the
host galaxy contributions -- are shown in \cref{fig:example_SED}.

\section{AGILE mock star catalog}
\label{sec:star}

Stars in the Milky Way and the Magellanic clouds are an important source of
contamination to consider for the photometric selection of \acp{AGN} and
quasars. For the stellar population in \ac{AGILE}, we use
``LSST SIM DR2'' \citep[][]{daltio2022ApJS..262...22D}, which is a
simulation of the \ac{LSST} stellar content down to
$r=27.5\,\mathrm{mag}$, including single and binary stars in the Milky Way and
in the Magellanic clouds.\footnote{%
  The simulations are available online at
  \url{https://datalab.noirlab.edu/lsst\_sim/index.php}.
}

These simulations are based on \texttt{TRILEGAL}
\citep{girardi2005A&A...436..895G}, a code for simulating the photometry of
resolved stellar populations in any Galaxy field -- as well as stellar systems
such as clusters and galaxies -- based on state-of-the art stellar evolution
tracks
\citep[][and references therein]{%
  bressan2012MNRAS.427..127B,%
  marigo2017ApJ...835...77M,%
  pastorelli2019MNRAS.485.5666P,%
  pastorelli2020MNRAS.498.3283P%
}.
For each star, the simulation provides astrometry and photometry, parallax, and
reddening, as well as physical and chemical properties, and proper motions.
Additionally, pulsation periods are provided for variability modeling for stars
in the \ac{CC} instability strip or that are \acp{LPV} in the
asymptotic giant branch (see~\cref{sec:variability}).

\citet[][]{daltio2022ApJS..262...22D} provide two separate stellar catalogs,
one for single stars (binary system fraction $f_\mathrm{bin} = 0.0$), and one
limited to binary systems ($f_\mathrm{bin} = 1.0$), thereby allowing users to
simulate any binary fraction by mixing the two catalogs. The binary star
catalog itself is composed of the physical properties of the individual stars
in the binary system, as well as the orbital parameters including the
inclination to the observer. This allows for the simulation of binary eclipses.
The provided binary star catalog only contains 10\% of the expected binary
systems. Therefore, following \citet[][Sect.~3.2]{daltio2022ApJS..262...22D} we
adopt $f_\mathrm{bin} = 0.40$ by first downsampling the single star catalog to
60\%, and add in four times the binary stars in the same region.

For the purpose of our simulations, for each star, we use the equatorial
coordinates, proper motions, and $ugrizy$ photometry. Additionally, we make use
of pulsation periods, when available, to simulate stellar variability, and for
binary systems we further use the orbital parameters in order to accurately
simulate binary eclipses. Both of these applications are described in
\cref{sec:variability}.

\section{The final AGILE mock truth catalog including AGNs, galaxies, and stars}
\label{sec:validation}

We use the aforementioned recipes in order to build a single mock truth catalog
of \acp{AGN}, galaxies and stars, on which we base the \ac{LSST} image
simulations as well as the photometric catalogs. This catalog spans a total
area of $24\,\mathrm{deg}^2$, centered on the coordinates of the COSMOS field.
We simulate the galaxy, \ac{AGN} and stellar populations as described, assuming
$0.2 < z < 5.5$ (imposed by the COSMOS2020 \ac{SMF}), and $\logMstar > 8.5$. We
populate each galaxy with $\lambda_\mathrm{SAR}$, label as \acp{AGN} the
population with $\loglambdaSAR > 32$, and derive the optical/\ac{UV} properties
for the \ac{AGN} population. Finally, we add in the stellar population directly
based on LSST SIM DR2, which is complete to $r < 27.5\,\mathrm{mag}$. The
$24\,\mathrm{deg}^2$ truth catalog contains
$\ntruthagnctn$ \ac{CTN} \acp{AGN}, $\ntruthagnctk$ \ac{CTK} \acp{AGN},
$\ntruthgalaxy$ non-active galaxies, and $\ntruthstar$ stars, to a grand total
of $\ntruth$ objects.

\subsection{Truth catalog contents}

The final truth catalog considered here contains all the columns from the
aforementioned processes for \acp{AGN}, galaxies, and stars. For \acp{AGN}, we
include the physical properties such as $L_\mathrm{X}$, optical/\ac{UV}
luminosities, $M_\mathrm{BH}$, $\lambda_\mathrm{SAR}$, \typei and \typeii
\ac{AGN} classification, \ac{AGN} $\EBV$, as well as the $ugrizy$ absolute and
apparent magnitudes derived from the \ac{AGN} \ac{SED}.

For galaxies (incl.\ AGN hosts), we include all the properties that
are provided by \textsc{EGG}. This includes right ascension, declination, $z$,
$M_\mathrm{star}$, \ac{SFR}, passive or star-forming classification,
morphological parameters (disk and bulge radii and luminosities), line
attenuation ($A_V$) for the galaxy bulge and disk, and $ugrizy$
absolute and apparent magnitudes derived from the galaxy \ac{SED}s. For more
details see the \textsc{EGG}
documentation\footnote{\url{https://cschreib.github.io/egg/files/EGG.pdf}} and
\cref{tab:catalog_columns}.

We note that the validation of the galaxy catalog and star catalog have already
been performed in their respective works
\citep{schreiber2017A&A...602A..96S,daltio2022ApJS..262...22D}. Here instead we
focus the remaining validation part on the \ac{AGN} catalog.

\subsection{Validation of the AGN mock catalog}

\subsubsection{X-ray luminosity function}

The first validation test performed on the mock truth catalog is to test the
\ac{AGN} \ac{XLF} against the literature values. In our approach, the \ac{AGN}
\ac{XLF} is directly set by the combination of the \ac{SMF} \ie galaxy number
density at a given $M_\mathrm{star}$, and \plambdafull. The \ac{XLF} thus
provides an important test on the completeness of the underlying X-ray \ac{AGN}
population. To compute the \ac{XLF}, we use the $24\,\mathrm{deg}^2$ truth
catalog and calculate directly the resulting \ac{XLF} at $\logLX > 42$ in
several $z$ bins. We show the results \ac{XLF} in \cref{fig:xlf}, and the
comparison to the compilation of literature values
\citep[see Appendix~A in][]{Shen:2020}.

\begin{figure*}
 \centering
 \includegraphics[width=1.0\linewidth]{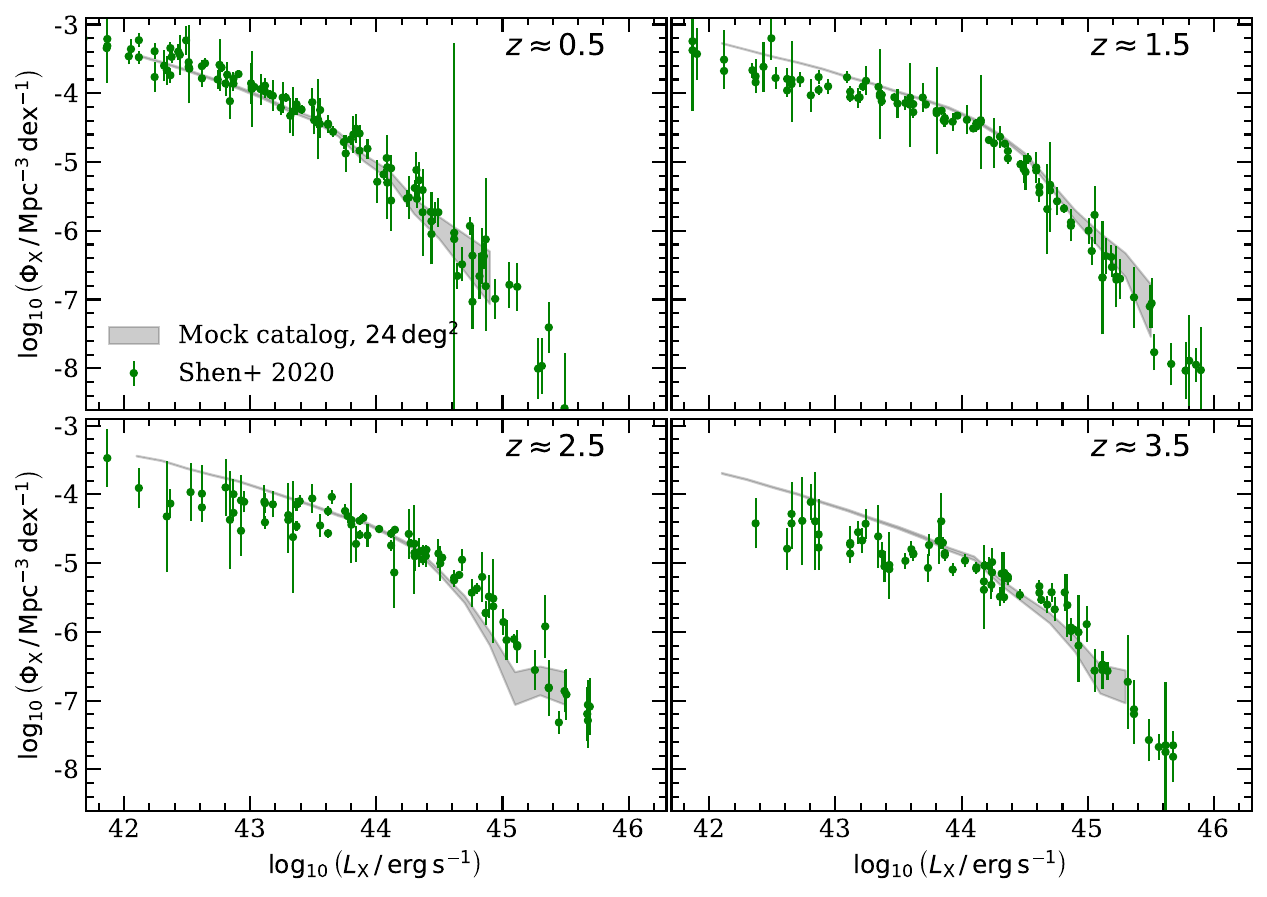}
 \caption{%
   \acl{XLF} from the mock catalog compared to the literature. The panels
   correspond to different $z$. The shaded region shows the \ac{XLF} from
   the $24\,\mathrm{deg}^2$ mock catalog, while the green markers show the
   observed \ac{XLF} compilation from various surveys \citep{Shen:2020}.
 }%
 \label{fig:xlf}
\end{figure*}

We find no apparent bias in the shape nor the normalization of the resulting
\ac{XLF} at $z = 0.5$--$3.5$, confirming that the X-ray \ac{AGN} population is
a complete sample of \ac{CTN} \acp{AGN}. This is unsurprising, as
\citetalias{Zou:2024} have already shown that their $p(\lambda_\mathrm{SAR})$
(regardless of quiescent or star-forming classification) combined with the
\citet{wright2018MNRAS.480.3491W} \ac{SMF} for all galaxies is consistent with
the \ac{XLF} of \citet{ueda2014ApJ...786..104U}. Therefore, we effectively
expand on their result by showing that a population of quiescent and
star-forming galaxies from the COSMOS2020 \ac{SMF} combined with their
$p(\lambda_\mathrm{SAR})$ for quiescent and star-forming galaxies are also in
agreement with the \ac{XLF} up to $z = 4$.

It is noteworthy that the recent discovery of ``little red dots''
poses a significant uncertainty for the luminosity function at $z > 4$
\citep[\eg][]{matthee2024ApJ...963..129M,ma2025ApJ...981..191M}. If their
debated nature turns out to be of \ac{AGN} origin, it would have a significant
impact on our work at $z > 4$. However, in the context of \ac{LSST}, the
majority of the \ac{AGN} population is still expected to reside at
intermediate $z \lesssim 4$, and detecting obscured \acp{AGN} in the first
place from optical/\ac{NIR} photometric data alone remains a challenge
\citep{euclid_bisigello2025arXiv250315323E,euclid_matamoro2025arXiv250315320E}.
The modular nature of \ac{AGILE} does allow for these populations to be
included in future versions.

\subsubsection{Optical number counts and luminosity function}

\begin{figure}
  \centering
  \includegraphics[width=\linewidth]{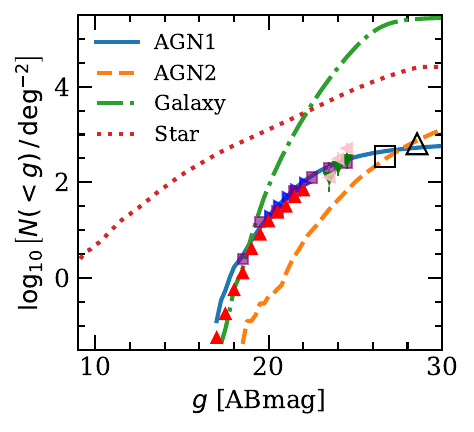}
  \caption{%
    Mock truth catalog ($24\,\mathrm{deg}^2$) $g$-band number counts for \typei
    and \typeii \acp{AGN}, galaxies, and stars. The colored markers show values
    from the literature \citep[][Fig.~10.7]{LSST-Science-Collaboration:2009}.
    The triangles correspond to \citet[][up]{Hartwick:1990}, \citet[][left and
    down]{Beck-Winchatz:2007}, and \citet[][right]{Richards:2006}, while the
    squares correspond to \citet{Wolf:2003}. The open triangle shows the
    \texttt{QSONumberCountMetric} for the COSMOS \acs{DDF} (see the text for
    the details), while the open square shows the expected QSO number counts
    for the ten-year \ac{LSST} survey \citep[][]{li2025arXiv251208654L}.
  }
  \label{fig:logn_B}
\end{figure}

\begin{figure*}
  \centering
  \includegraphics[width=\linewidth]{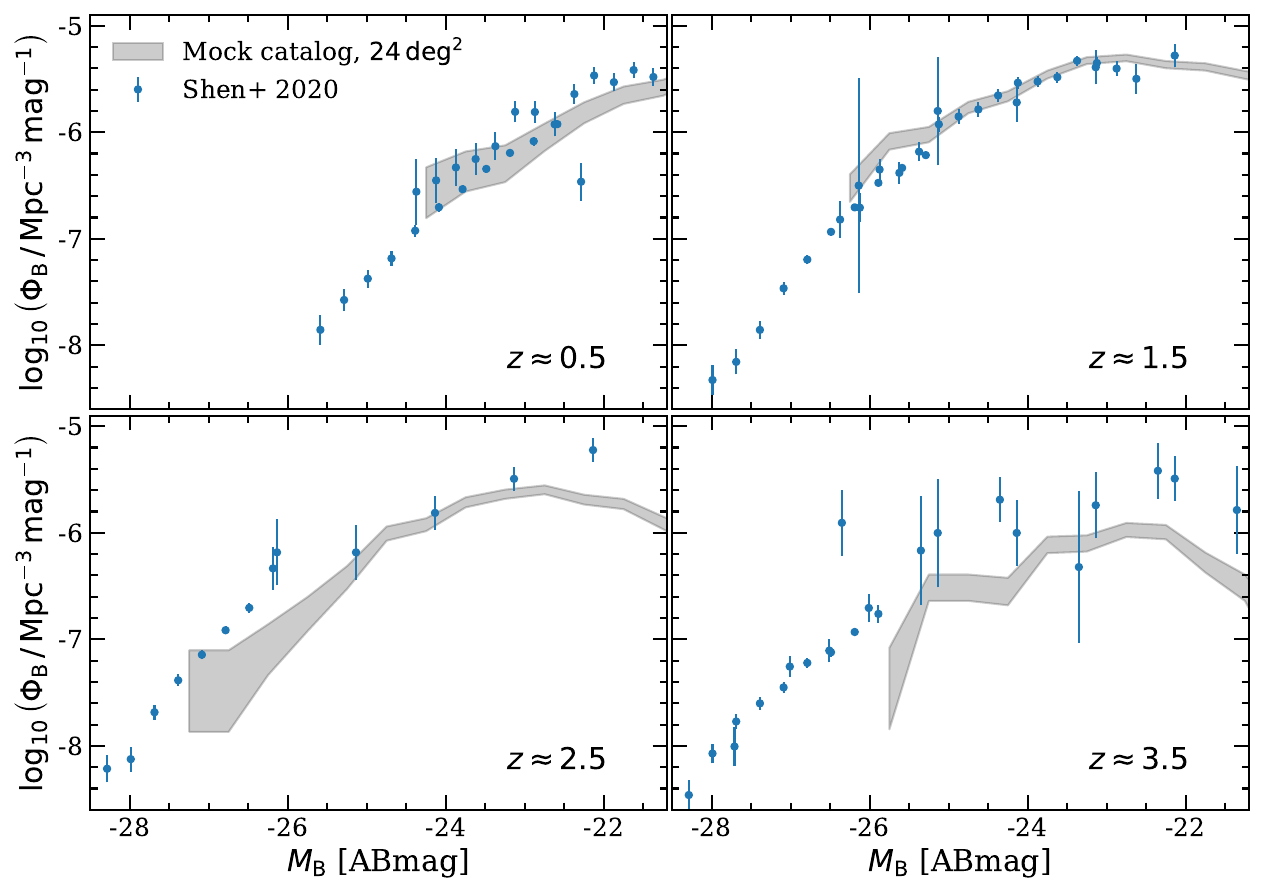}
  \caption{%
    The $B$-band quasar luminosity function. The shaded regions show the
    luminosity function of mock \typei \ac{AGN} from the $24\,\mathrm{deg}^2$
    truth catalog. The markers show the recent data compilation of the quasar
    luminosity function from \citet{Shen:2020} at the redshift indicated by the
    text.
  }%
  \label{fig:qlf}
\end{figure*}

The optical \ac{AGN} and quasar populations arise in the truth catalog from the
combination of X-ray and optical \ac{AGN} properties. That is, the expected
number of quasar-like optical \acp{AGN} is mainly driven by the \ac{XLF},
$\lx$--$\luv$ relation \citep{Lusso:2010}, and the \typeii \ac{AGN} fraction
\citep{Merloni:2014}. We have used the mock truth catalog to
directly calculate both the expected sky number density of optical quasars, as
well as the quasar luminosity function. We show the $g$-band number counts for
the various classes of objects in \cref{fig:logn_B}, and the $B$-band quasar
luminosity function in \cref{fig:qlf}.

We find that the $g$-band number counts as computed from the
$24\,\mathrm{deg}^2$ truth catalog are overall consistent with the literature
\citep{Richards:2006,Wolf:2003,Hartwick:1990,Beck-Winchatz:2007}
at $g<25$. However, we find a slight overproduction of quasars at the bright
end $g<18$, likely ascribed to the choice of extrapolation of the \typeii
\ac{AGN} fraction (see below). Moreover, within the $24\,\mathrm{deg}^2$ truth
catalog, the errors in the number counts remain still relatively large,
compared to the all-sky surveys from which the bright-end $g<18$ observed
number counts are measured from. At the expected \ac{LSST} COSMOS ten-year
depth, we also find the mock \ac{AGN} number counts to be consistent with the
expected \ac{LSST} \ac{QSO} number counts. Specifically, we refer to the
\texttt{DD:COSMOS CoaddM5} and \texttt{DD:COSMOS QSONumberCountsMetric} metrics
calculated within the \ac{LSST} \acl{MAF} \citep[\acs{MAF}\acused{MAF}; ][and
\cref{sec:image}]{jones2014SPIE.9149E..0BJ}, and the expected QSO number counts
from the ten-year \ac{LSST} survey \citep[black
squares;][]{li2025arXiv251208654L}. Assuming the ten-year COSMOS \ac{DDF} depth
of $g=28.56$, we find ($530 + 754)\,\mathrm{deg}^{-2}$ \typei and \typeii
\acp{AGN} in the mock. In the meanwhile, \texttt{DD:COSMOS
QSONumberCountMetric} (baseline v4.0) suggests $637\,\mathrm{deg}^{-2}$,
assuming a dithered COSMOS \ac{DDF} area of ${\sim} 2 \times
9.6\,\mathrm{deg}^2$. We note that here the $g$-band flux refers solely to the
\ac{AGN} flux, ignoring any host-galaxy contribution, which observationally
would further complicate accurate measurements of the faint end of the number
counts.

In addition, we measure the optical luminosity function of the mock \typei
\ac{AGN} population and compare it to the literature across a wide range in $z
\approx 0.5$--$4.0$ (\cref{fig:qlf}). We note that the luminosity function was
computed using the \ac{AGN} intrinsic absolute magnitude that was de-reddened
according to the $\EBV$. We find that both the shape and the normalization of
the $B$-band quasar luminosity function are in general agreement with the
observed one \citep{Shen:2020}. At higher $z \approx 3.5$ the mock
under-predicts the luminosity function at the ${\sim} 0.2$--$0.4\,\mathrm{dex}$
level in the $M_B > -25$ regime. As anticipated, these discrepancies could be
mitigated by fine-tuning the extrapolation strategy of the \typeii AGN fraction
\citep{Merloni:2014}, especially at $z \sim 0.5$ and $z > 3$. Moreover, the
low-luminosity tail of the $z \sim 3.5$ luminosity function suffers from more
uncertain comparison data. Given these, we refrain from further optimizing the
extrapolation of the \typeii \ac{AGN} fraction.

\subsubsection{Active black hole mass function}

\begin{figure}
  \centering
  \includegraphics[width=.9\linewidth]{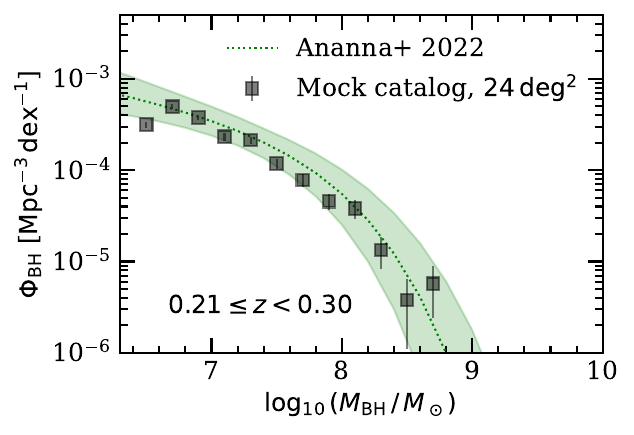}
  \caption{%
      Local active \ac{BHMF}. The mock truth catalog (symbols) and observed
      \citep[lines and shaded region; ][]{ananna2022ApJS..261....9A} active
      black hole mass functions both correspond to Eddington ratios $-3 <
      \logten \lambda < 1$. The mock $\lambda$ is estimated from
      $M_\mathrm{BH}$ and $L_\mathrm{X}$, assuming a constant bolometric
      correction $L_\mathrm{X} \,/\, L_\mathrm{bol} = 25$.
  }
  \label{fig:bhmf}
\end{figure}

The active \ac{BHMF} is an important observable validating the assignment of
\acp{AGN} to galaxies. It is the by-product of three key components in the mock
truth catalogs, \ie the galaxy \ac{SMF} \citep{weaver2023A&A...677A.184W},
the $M_\mathrm{BH}$--$M_\mathrm{star}$ relation \citep[\eg][and
\cref{sec:mbh}]{zou2024ApJ...976....6Z},
and the probability of a black hole being active \citepalias[\eg][]{Zou:2024}.
In order to further validate the mock truth catalog, we compare the predicted
active \ac{BHMF} from the $24\,\mathrm{deg}^2$ mock catalog to the observed
local ($0.21 \leq z < 0.30$) active \ac{BHMF} of
\citet{ananna2022ApJS..261....9A}.
Their analysis is based on a complete $z < 0.3$
sample of ultra-hard selected ($14$--$195\,\mathrm{keV}$) X-ray
\acp{AGN} from the Swift-BAT \ac{AGN} Spectroscopic Survey \citep[BASS;
][]{koss2017ApJ...850...74K,koss2022ApJS..261....1K},
which is more sensitive towards \ac{CTK} \acp{AGN} compared to the
typical $2$--$10\,\mathrm{keV}$ band selection.
Thus, we calculate the active \ac{BHMF} including both \ac{CTN} and \ac{CTK}
\acp{AGN}, and Eddington ratio
$\lambda_\mathrm{Edd}$ limits as in \citet{ananna2022ApJS..261....9A}.
To be consistent with the assumptions in the continuity equation
(\cref{sec:mbh}), we convert mock ${M_\mathrm{BH}}$ and ${L_\mathrm{X}}$ to
$\lambda_\mathrm{Edd}$ assuming a constant bolometric correction
$k_\mathrm{bol} = 25$. As shown in \cref{fig:bhmf}, the mock truth catalog
successfully reproduces the normalization and general shape of the local active
\ac{BHMF}.

The comparison with higher-$z$ \ac{BHMF} measurements is less conclusive with
the mock \ac{BHMF} systematically below the best-fit model by e.g.
\citet{schulze15}. This is unsurprising given the large uncertainties in the
$M_\mathrm{BH}$--$M_\mathrm{star}$ relation at high $z$. While the continuity
equation adopted here predicts a nearly $z$-invariant
$M_\mathrm{BH}$--$M_\mathrm{star}$ relation consistent with the local relation
\citep{reines15}, observational constraints at higher-$z$ remain highly
uncertain and span a wide range of normalizations
\citep[\eg][]{Merloni:2010,suh20,pacucci23,maiolino2024A&A...691A.145M,tanaka25}

\section{Optical variability}
\label{sec:variability}

The mock catalog created thus far depicts a purely static universe. However,
the strength of \ac{LSST} lies in its high cadence, where a sky position will be imaged on average once every three nights (see \cref{sec:image}).
Here we describe the recipes to add in optical variability into the
simulation for different classes of objects. We first include a model for
\ac{AGN} variability, and second we consider stellar variability for \acp{CC},
\acp{LPV}, and binary star systems. We do not consider transient phenomena such
as supernovae, or tidal disruption events.

\subsection{AGN variability}
\label{sec:agnvariability}

Active galactic nuclei exhibit stochastic variability at all wavelengths and
with timescales ranging from minutes to years
\citep{paolillo2025NCimR..48..537P}.
Continuum optical light curves have been empirically described as a \acl{DRW}
\citep[\acs{DRW}\acused{DRW}; ][]{Kelly2009}, a first-order continuous
autoregressive moving-average (CARMA) process modeling variability with two
main parameters: the amplitude of correlation decay $\sigma$, and a
characteristic damping timescale $\tau$. In detail, variability is described
with an exponential decay \ac{ACF}, where the covariance between two points
separated by $\Delta t$ is given by a combination of Gaussian processes as:
\begin{equation}
    K(\Delta t) = \sigma^2\,\mathrm{ACF}(\Delta t) =
    \sigma^2 \, \exp \left( -\Delta t \,/\, \tau \right).
    \label{eq:drw_corr}
\end{equation}
The \ac{DRW} predicts a power-law power spectral density for \acp{AGN}, with a
spectral index of $-2$, flattening to zero for $\gg \tau$. Such a model has
been successful in describing both the stochastic and the typical red-noise
trend of \ac{AGN} light curves with lengths of the order of a few years,
although deviations have been observed on both longer and shorter timescales
with possibly steeper high-frequency slopes and longer decorrelation timescales
for a complete flattening \citep[\eg][]{Mushotzky+11,Guo+17,Arevalo2024}.
Albeit more refined methods have been proposed, such as fitting light curves
with higher-order CARMA processes (\eg the damped harmonic oscillator,
\citealt{Yu2022}), direct or indirect estimation of the power spectral density
\citep[\eg][]{Kelly2014,Petrecca2024}, or unsupervised machine learning
analysis of time series (\citealt{Tachibana2020}), there is still no
definitive model to describe \ac{AGN} optical variability.

While it might not be the most accurate description on short (less than a few
days) or very long (more than a few years) timescales, the \ac{DRW} provides a
robust and well-tested first-order
approximation and comes with a flexible and easy way to produce simulated light
curves \citep{Suberlak2021}. Thus it provides a controlled baseline model with
which the other aforementioned models may be compared to. Moreover, it has a
direct connection with the \ac{SF}, which is typically used to parametrize
\ac{AGN} light curves in the temporal domain instead of the power spectrum. The
\ac{SF} is defined as the root mean square magnitude difference as a function
of the time difference $\Delta t$ between observation pairs
(\citealt{Kozlowski2016}), and it is related to the \ac{ACF} (and thus the
\ac{DRW}) as
$\mathrm{SF}(\Delta t)
= \sqrt{2\,\sigma^2 \left[ 1 - \mathrm{ACF}(\Delta t) \right]}
= \mathrm{SF}_\infty \sqrt{1 - \exp \left(-|\Delta t| \,/\, \tau \right)}$.
Here, we used the \ac{ACF} of the \ac{DRW} from \cref{eq:drw_corr},
and introduced $\mathrm{SF}_\infty = \sqrt{2}\,\sigma$ as the variability
amplitude for $\Delta t \rightarrow \infty$, typically used as a parameter in
\ac{ACF} analyses together with $\tau$.

Both \ac{DRW} and the associated \ac{SF} parameters have been observed to
correlate with physical \ac{AGN} properties, such as $M_\mathrm{BH}$, accretion
rate, bolometric luminosity and rest-frame wavelength
\citep[\eg][]{MacLeod2010,Zu+13,Kasliwal+15,Suberlak2021}. In particular, we
use the latest \ac{DRW} parameters calibrated using the $15\,\mathrm{yr}$
baseline for $9248$ quasars selected from the SDSS Stripe-82 and cross-matched
with Pan-STARRS1 (\citealt{Suberlak2021}, references therein, and Table~1 of
\citealt{paolillo2025NCimR..48..537P}) in order to assign each \ac{AGN} in the
simulation a ten-year \ac{DRW} light curve. The parameters $\tau$ and
$\mathrm{SF}_\infty$ scale with the physical \ac{AGN} properties via
\begin{equation}
\begin{split}
  \logten f &= A + B \logten (\lambda_\mathrm{RF} \,/\, 4000\,{\angstrom}) \\
            &+ C (M_i + 23) + D \logten (M_\mathrm{BH} \,/\, 10^9 M_\odot),
  \label{eq:logf}
\end{split}
\end{equation}
where $f$ is used to denote either $\tau$ or $\mathrm{SF}_\infty$, and both
parameters have their separate best-fit values $A$, $B$, $C$, and $D$ as
summarized in \citet[][Table~2]{Suberlak2021}.
Here $\lambda_\mathrm{RF}$ and $M_i$ correspond to the rest-frame wavelength
(set by the \ac{LSST} band and $z$) and the absolute magnitude in the $i$-band,
respectively. Light curves are simulated using the \ac{DRW} implementation by
\citet{Kovacevic2021}, and providing the $\mathrm{SF}_\infty$ and $\tau$
parameters derived from the scaling relations as input.

The scaling relations to calibrate the \ac{DRW} parameters used for this
simulation were derived from a sample of quasar-like \acp{AGN} with a certain
range of masses and bolometric luminosities. Although this is quite large, with
$\logMBH \approx 7.0$--$10.5$ and $\logLbol \approx 44.5$--$47.5$, it does not
cover the full variety of simulated \acp{AGN}. Whenever we had to simulate a
\ac{DRW} for a source outside the range of SDSS Stripe-82 quasars, we
extrapolated the scaling relations and confirmed that the recovered light
curves had variability amplitudes compatible to those reported in literature.

Another important consideration is related to \typeii \acp{AGN}, which are
typically missed by optical surveys because of the high degree of obscuration.
For this reason, large statistical studies of variability usually focus on
non-obscured sources. However, this is expected to change with \ac{LSST} where
subtracting the host-galaxy contribution via \ac{DIA} on the entire dataset
with an extended temporal baseline is possible. Many recent works report
\typeii \acp{AGN} to feature suppressed variability with a flatter
\ac{SF} \citep[\eg][]{De-Cicco:2022,Lopez-Navas2023}. To simulate \typeii
\ac{AGN} light curves compatible with these observations, we used the same
scaling relations as for \typei{}, but adding an over-damping factor of ten to
both $\mathrm{SF}_\infty$ and $\tau$ \citep{Lopez-Navas2023}. Although this
empirical factor introduces some uncertainty, it is worth stressing that the
main aim of this simulation is to test the capability of \ac{LSST} to recover
the input, focusing on any bias due to the observing strategy, the photometry,
or the reduction pipeline.

Here we have only focused on the average red-noise variability of the typical
\acp{AGN}. For example, our \ac{AGN} variability model does not consider the
time lag between the bands, but there are specific works studying this
in the context of \ac{LSST} \citep[\eg][]{czerny2023A&A...675A.163C}. Also,
other relatively rare sources such as blazars, binary \acp{AGN}, or extreme
variability phenomena (\eg changing-look \acp{AGN}, outbursts, and deep fades;
see \citealt{komossa2026AdSpR..77.4041K} for a recent review) are not included.

\subsection{Star variability}

Using the orbital parameters from the binary star catalog (\cref{sec:star}), it
is straightforward to simulate the light curves of eclipsing binaries in the
\ac{LSST} $ugrizy$ filters. Toward this goal, we have used the \texttt{batman}
software \citep{kreidberg2015PASP..127.1161K}. Further stellar variability
relevant for this study comes from stellar pulsation. In its current version,
the \texttt{TRILEGAL} simulation includes only two types of pulsating stars,
namely \acp{CC} and \acp{LPV}, for which pulsation periods are provided. Other
variability parameters for these stars (amplitudes, light curves), as well as
additional pulsating star types are planned for future versions of the
simulation \citep{daltio2022ApJS..262...22D}.
Considering the scope of the present work and the effort that would be required
to add information for other variability types, we preferred to limit our study
to the most common variability manifestations that have already been validated
in the simulations. For these, we adopt the following procedures in order to
simulate light curves.

\subsubsection{Classical Cepheids}
\label{sec:lc_cc}

The \ac{CC} pulsation models used to generate the theoretical light curves in
the \ac{LSST} filters are derived from a nonlinear, convective hydrodynamical
approach, specifically employing the \texttt{Stellingwerf} hydrodynamical code
\citep[][]{bono2000ApJ...543..955B,bono2000ApJ...529..293B}. This dataset
extends previous pulsation models \citep[\eg][and references
therein]{bono2000A&A...354..610C,fiorentino2007A&A...476..863F} by including,
simultaneously, variations in chemical composition, mass-luminosity
($M$--$L$) relation,
and superadiabatic convection efficiency. The models are computed for
fundamental (F), first overtone (FO), and second overtone (SO) pulsation modes,
considering four different chemical compositions: Low-metallicity cases:
$Z=0.004$, $Y=0.25$ and $Z=0.008$, $Y=0.25$, Solar-like metallicity: $Z=0.02$,
$Y=0.28$ and supersolar metallicity: $Z=0.03$, $Y=0.28$. The models cover a
wide range of stellar parameters, including effective temperature:
$3600\,\mathrm{K}$ to $7200\,\mathrm{K}$, in steps of $100\,\mathrm{K}$ and
mass range from $3$ to $11\,M_{\odot}$, in steps of $1\,M_{\odot}$. The
convective efficiency is parameterized using the mixing length parameter
$\alpha_\mathrm{ml} = 1.5$ (standard convective efficiency, $\alpha_\mathrm{ml}
= 1.7$ and $\alpha_\mathrm{ml} = 1.9$).

Moreover, three $M$--$L$ relations are adopted, following the formulation
provided by \citet[][]{bono2000ApJ...543..955B}: case A \ie canonical models
with no core overshooting, rotation, or mass loss, case B \ie non-canonical
models with a moderate luminosity increase of $\Delta\logten(L\,/\,L_\odot) =
0.2\,\mathrm{dex}$ and case C \ie non-canonical models with a stronger
luminosity increase of $\Delta\logten(L\,/\,L_\odot) = 0.4\,\mathrm{dex}$. For
this study, we selected models that satisfy the luminosity range adopted in the
simulation ($2.5 < \logten (L\,/\,L_\odot) < 4.8$) specifically:
\begin{itemize}
  \item Fundamental and first overtone mode models,
  \item all four chemical compositions,
  \item $\alpha_{ml} = 1.5$,
  \item non-canonical $M$--$L$ relation (case B, over-luminous by $0.2\,\mathrm{dex}$).
\end{itemize}
For a comprehensive description of the assumptions underlying the \ac{CC}
pulsation models, the key results such as the topology of the instability
strip, the period-luminosity-color (PLC) relations, and \ac{LSST} light curves
adopted for this work, we refer the reader to
\citet[][]{desomma2020ApJS..247...30D,desomma2022ApJS..262...25D,desomma2024MNRAS.528.6637D}.

\subsubsection{Long-period variables}
\label{sec:lc_lpv}

Long-period variable stars represent the final stages in the evolution
of low- to intermediate-mass stars ($0.8\lesssim
M_\mathrm{init}\,/\,\mathrm{M}_{\odot}\lesssim 8$). They are cool, red giant
stars with peak emission in the red or \ac{NIR}, and display photometric
variability with periods of order of several days to a few years, often with
cycle-to-cycle variations and various degrees of regularity. Depending on the
spectral range of observation and intrinsic stellar properties, they span a
range of photometric amplitudes, from a milli-magnitude level up to more than
ten magnitudes in the visual filters.

The \texttt{TRILEGAL} simulation of the \ac{LSST} stellar content includes pulsation
periods of \acp{LPV} based on stellar pulsation models of
\citet{trabucchi2019MNRAS.482..929T,trabucchi2021MNRAS.500.1575T}, as well as
an indication of the dominant pulsation period. Synthetic light curves of \acp{LPV},
and their photometric amplitudes, are not currently available in \texttt{TRILEGAL}
simulations. To the best of our knowledge, no prescription is available from
the modern scientific literature to predict these features as a function of
global stellar parameters. Therefore, we rely on a simplified, semi-empirical
description.

We take advantage of the fact that \acp{LPV} follow a period-amplitude relation
\citep[\eg][and references therein]{trabucchi2019MNRAS.482..929T}. We examined
the $I$-band photometric time series of \acp{LPV} in the Magellanic Clouds
observed as part of the Optical Gravitational Lensing Experiment \citep[OGLE;
][]{soszynski2009AcA....59..239S,soszynski2011AcA....61..217S}, and derived an
approximate analytic relation
\begin{equation}\label{eq:fit_ogle_period_amplitude}
  \logten \sigma_I \simeq
  1.5\,\logten \left( P_1 \,/\,\mathrm{day} \right) - 4.0,
\end{equation}
where $P_1$ is the primary variability period and $\sigma_I$ is the standard
deviation of the $I$-band light curve, which is a tracer of the variability
amplitude. To convert it into the \ac{LSST} $ugrizy$ filters we adopt the
results of \citet{iwanek2021ApJS..257...23I}. Finally, we assume that the light
curve can be modeled as a simple sine with randomized phase offset.

This approach to model the photometric variability of \acp{LPV} is rather
crude: the period-age relation suffers from a relatively large scatter, and
both the regularity of the light curve and the relation between peak-to-peak
amplitude and the standard deviation depend on the degree of multi-periodicity
of a star. Nonetheless, it provides us with an efficient method of estimating
the order-of-magnitude impact of \acp{LPV} as a source of confusion against
\ac{AGN} variability, which is an acceptable trade-off for the purpose of the
present paper.

\section{Image simulations}
\label{sec:image}

\begin{figure*}
  \includegraphics[width=\linewidth]{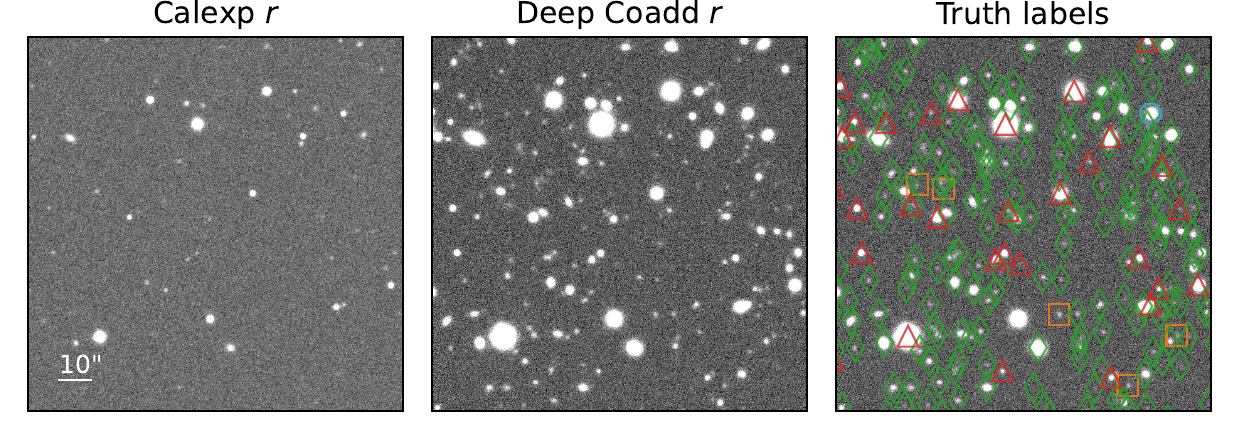}
  \caption{%
    Example single-visit (\texttt{calexp}, left panel), deep coadd (middle
    panel) images, and the underlying truth catalog labels (right panel). The
    images correspond to the $r$-band with a small \ac{FOV} of $2\arcmin$
    centered around $\mathrm{RA} = 150\degf2121877$, $\mathrm{dec} =
    2\degf1904921$ (J2000), while the LSSTCam has a single detector side length
    of ${\sim} 13\arcminf5$. The $r$-band image corresponds to a stack of
    $N=289$ individual $29.2\,\mathrm{s}$ exposures. In the right panel, the
    colors and symbols mark the positions of \typei \acp{AGN} (blue circles,
    one bright source in the top-right), \typeii \acp{AGN} (orange squares),
    galaxies (green diamonds), and stars (red triangles). It is noteworthy that
    one bright $r = 19.05$ star is missing its truth label (right panel,
    bottom-center) due to the contamination of a fainter $r = 24.65, z=1.62$
    galaxy at a separation of $1\arcsecf7$.
  }
    \label{fig:calexp_coadd_truth}
\end{figure*}

The detailed \ac{LSST} ten-year survey plan is defined in the baseline strategy
\citep[see][and references
therein]{ivezic2019ApJ...873..111I,bianco2022ApJS..258....1B}. Overall, the
majority of the survey time is expected to be dedicated to the \ac{WFD} which
covers the southern sky, while some $7\%$ of survey time is dedicated to the
\acp{DDF}. The \acp{DDF} are five extragalactic fields covered by one to two
telescope pointings and to several magnitudes deeper compared to the
\ac{WFD}.\footnote{%
  The final depth is only one of the many available survey metrics that are
  available online: \url{https://survey-strategy.lsst.io/}.
}
Notably, the COSMOS \ac{DDF} is the deepest of the planned \acp{DDF}, and
receives approximately twice the number of visits compared to any other
\ac{DDF}. On the contrary, the \Euclid Deep Field South visits are split
between two \ac{LSST} pointings.

The \ac{LSST} survey strategy is described as a sequence of visits, each
corresponding to a single telescope pointing, a chosen filter ($ugrizy$), and
the exposure time (nominally $t_{grizy} \sim 29.2\,\mathrm{s}$, and $t_u =
38\,\mathrm{s}$).
Within \ac{LSST}, survey strategies are evaluated and simulated using the
\acl{OpSim}\footnote{\url{https://rubin-sim.lsst.io/}}
(\acs{OpSim}\acused{OpSim}),
and the \acl{MAF} \citep[\acs{MAF}; ][]{jones2014SPIE.9149E..0BJ}. In
addition to the telescope information, these simulations also provide the
expected local sky conditions such as the positions of the Sun and the Moon,
air mass, and seeing. Here we adopt the latest survey strategy (at the time
of initiating this work), ``baseline v4.0'' as the reference.
Differences between various survey strategies are detailed on-line
.\footnote{\url{https://survey-strategy.lsst.io/baseline/changes.html}}
We focused on the COSMOS \ac{DDF}, which is one of the most well-known regions
of the extragalactic sky in terms of depth and wavelength coverage
\citep[see][]{weaver2022ApJS..258...11W}.

According to baseline v4.0, the ten-year \ac{LSST} survey contains a total of
$2\,038\,634$ visits, with $43\,594$ in the COSMOS \ac{DDF}. The total number
of COSMOS visits per \ac{LSST} $ugrizy$ band are $2185, 5033, 9919, 9994,
11\,538, 4925$, respectively. Temporally, the first half of the visits is
completed within the first three years of the survey. In the \acp{DDF}, visits
are further grouped into ``sequences'', so that every time a \ac{DDF} is being
observed, multiple exposures are taken in succession instead of a single one
($N_{ugrizy} = 8$, $10$, $20$, $20$, $24$, $18$). In contrast, in baseline v5.0
(the latest strategy at the time of writing) the ``ocean'' \ac{DDF} strategy is
adopted. Here, \ac{DDF} sequences would alternate between shallow (less visits
per sequence) and deep (more visits per sequence) seasons. The most of the
\ac{DDF} visits would then occur in the deep seasons, of which there is at
least one per \ac{DDF}. These \acs{OpSim} runs are available
on-line.\footnote{\url{https://usdf-maf.slac.stanford.edu/}}

\subsection{The instance catalogs}

The \ac{LSST} cadence is used by \ac{AGILE} -- together with the original mock
catalog and light curve information -- to generate the so-called instance
catalogs. These catalogs correspond to a snapshot of the mock catalog at the
time of each \ac{LSST} visit. The instance catalogs describe the position
(varying for stars based on their proper motions), apparent magnitude (varying
for stars and \acp{AGN} based on their light curves), and the morphology of
each object. Following \citet[][Sect.~3.2]{schreiber2017A&A...602A..96S},
we describe galaxies as a combination of two Sérsic profiles
assuming $n=4$ for the bulge and $n=1$ for the disk. For \acp{AGN}, an
additional point-like component is added on top of the galaxy bulge and disk
at the center, while stars are considered to be point-like sources.

\subsection{Raw simulated LSST images}

The synthetic \ac{LSST} raw images are generated by \ac{AGILE} from the
instance catalogs using \ac{DESC} software called
\textsc{ImSim}.\footnote{\url{https://github.com/LSSTDESC/Imsim}} This software
simulates the $3.2\,\mathrm{Gigapixel}$ LSSTCam instrument
accurately,\footnote{\url{http://lsstdesc.org/imSim/features.html}} including
electronic (\eg bias, dark current, and non-uniformity) and atmospheric effects
(\eg seeing). The LSSTCam is a large mosaic camera, composed of a total of
$189$ individual detectors, which fully cover the $9.6\,\mathrm{deg}^2$
\ac{FOV} of the telescope. With \textsc{ImSim}, any subset of the $189$
detectors may be simulated individually, resulting into realistic synthetic
\ac{LSST} raw exposures.

\subsection{Final simulated science images}

Raw LSSTCam exposures then converted into calibrated single-visit
exposures called \texttt{calexp}\footnote{%
  In \ac{LSST} \acl{DP1}, \texttt{calexp} images have been renamed to
  \texttt{Visit images} (\url{https://doi.org/10.71929/rubin/2570311}). A
  single \texttt{calexp} image acts as a container for the science, variance,
  and mask (recording processing status or issues) images.
}
images by first removing the signal of the instrument and performing the
calibration in both astrometry and photometry. For astrometric and photometric
calibration, the mock truth catalog is used as the reference catalog, including
proper motions for stars. We do not consider photometric nor astrometric errors
arising from the reference catalog, which is an additional source of error in
the actual \ac{LSST} data. As per the \ac{LSST} requirements, the
expected \texttt{calexp} depths in the $ugrizy$ bands are $23.9, 25.0, 24.7,
24.0, 23.3, 22.1$ ($5\,\sigma$), respectively
\citep{bianco2022ApJS..258....1B}. However, individual \texttt{calexp} image
depths will depend on the survey cadence and will be further modified by air
mass, seeing, and the final survey strategy.

Finally, the individual single-visit \texttt{calexp} images are combined to
create deep coadded images in each of the \ac{LSST} $ugrizy$ bands. In the
coaddition, the sky is first tessellated into tracts, which are defined as
partially overlapping rectangular regions with a side length of
$1.6\,\deg$. Each tract is divided further into $7\,\times\,7$
patches, while each patch is further subdivided into $4100\,\times\,4100$
pixels with a pixel scale of $0\arcsecf2$. In each pixel, the coadded value is
the result of resampling and combining the corresponding single-visit
\texttt{calexp} images. In the \ac{LSST} requirements, the expected ten-year
coadded image depths in $ugrizy$ are $26.1$, $27.4$, $27.5$, $26.8$, $26.1$,
and $24.9$ ($5\,\sigma$), respectively \citep{bianco2022ApJS..258....1B}.

\cref{fig:calexp_coadd_truth} shows an example zoom-in of \texttt{calexp} and
coadded images in the $r$-band. The coadded image includes a total of $289$
visits in the COSMOS field based on the baseline v4.0 and the
$24\,\mathrm{deg}^2$ truth catalog, highlighting the increase in depth
incrementally as the survey progresses. In addition,
\cref{fig:calexp_coadd_truth} shows the positions of individual truth catalog
\acp{AGN}, galaxies, and stars.

For these tasks of reduction, calibration, and coaddition, \ac{AGILE} uses the
\ac{LSST} Science pipelines \citep[v8.0.0,
w\_2024\_16;][]{bosch2018PASJ...70S...5B}, For more details, also refer to
\citet{lsst2021ApJS..253...31L}.

\section{Photometric catalogs}
\label{sec:pipeline}

The last step of \ac{AGILE} is to perform the photometric analysis of the
single-visit and coadded images. This is done using the \ac{LSST} Science
pipelines (v8.0.0, w\_2024\_16). Starting with the deep coadded images,
\ac{AGILE} performs source detection at a significance level of $\SNR > 5$ in
order to create the object table. For each detection, the object table contains
the set of $ugrizy$ positions and flux measurements, the characterization of
the morphology of the source, and quality flags. \ac{AGILE} also performs
forced photometry at the coordinates of each object in each single-visit
\texttt{calexp} image. This results in a forced photometry catalog, where each
object has their corresponding measured $ugrizy$ light curve in accordance with
the input \ac{LSST} baseline. These data products are detailed in the
Rubin-LSST documentation to which we provide the relevant links in
\cref{app:columns}.

Some of the Rubin Data Products have different scientific applications compared
to ours, such as measurements from the \ac{DIA}. Having the focus on the
\ac{AGN} population, we proceed on with the discussion of the photometry within
the object catalogs from the deep coadded images, as well as the forced
photometry catalogs from the single-visit images.

Throughout the following sections, we base the analysis on the \ac{AGILE} DR1
photometric catalog corresponding to $1\,\mathrm{deg^2}$ ($21$ out of $189$
LSSTCam detectors) and $\nvisitdri$ visits over three years of observations in
the COSMOS \ac{DDF}. The catalog and the assumed detector layout are described
further in \cref{app:dr1}.

\subsection{Object catalogs}

The object catalog is constructed by measuring the properties of extracted
sources ($\SNR > 5$) from the coadded $ugrizy$ images. Each object defined this
way contains a total of $1186$ columns of information consisting of astrometric
(\eg RA, Dec, and their respective errors), morphological (\eg Kron radii and
extendedness), and photometric (\eg $ugrizy$ flux measurements) measurements.
The fluxes and their errors are provided for a wealth of definitions, including
aperture, \ac{PSF}, and composite model or \texttt{cModel}. The fluxes and
their quality flags are also reported for each band separately.

As an example of the source density, we use the object catalog produced from
the $\nvisitdri$ visits in COSMOS\@. We select a single patch which corresponds
approximately to the area of one LSSTCam detector ${\sim} 0.05\,\mathrm{deg^2}$
with a total of $12\,082$ objects. Out of these, $9171$ have $\SNR > 5$
(\texttt{psfFlux}) in the $r$-band, and a further $1985$ have also $r<24$,
giving an approximate source density of $0.19\,\mathrm{M}\,\mathrm{deg}^{-2}$
with $\SNR > 5$, and $40\,\mathrm{K}\,\mathrm{deg}^{-2}$ with $\SNR > 5$ and $r
< 24$. These densities may be directly compared to the underlying truth
catalog, where the $24\,\mathrm{deg^2}$ truth catalog suggests total (\ac{AGN},
galaxy, or star) source densities of $50\,\mathrm{K}\,\mathrm{deg}^{-2}$
($40\,\mathrm{K}\,\mathrm{deg}^{-2}$) at $r<24$ ($r<23.7$).
The apparent discrepancy arises from a combination of deblending losses and
underestimated galaxy true fluxes by the \texttt{psfFlux}. Indeed, by computing
the number of close angular pairs in the truth catalog in a similarly-sized
area and assuming $r < 24$, we find a significant incompleteness in terms of
angular pairs at $<2\arcsec$ (close to $100\%$ at $<1\arcsec$). However, this
accounts only for a few hundred missing pairs overall and therefore cannot
explain the full discrepancy. At larger separation, the discrepancy is
instead consistent with underestimated galaxy fluxes by \texttt{psfFlux}.
Repeating the density calculation with \texttt{cModelFlux} yields a source
density of $51\,\mathrm{K}\,\mathrm{deg}^{-2}$, supporting this idea. This is
examined in more detail in the next section.%

\subsection{Photometric Accuracy}
\label{sec:photometric_accuracy}

\begin{figure*}
    \includegraphics[width=\linewidth]{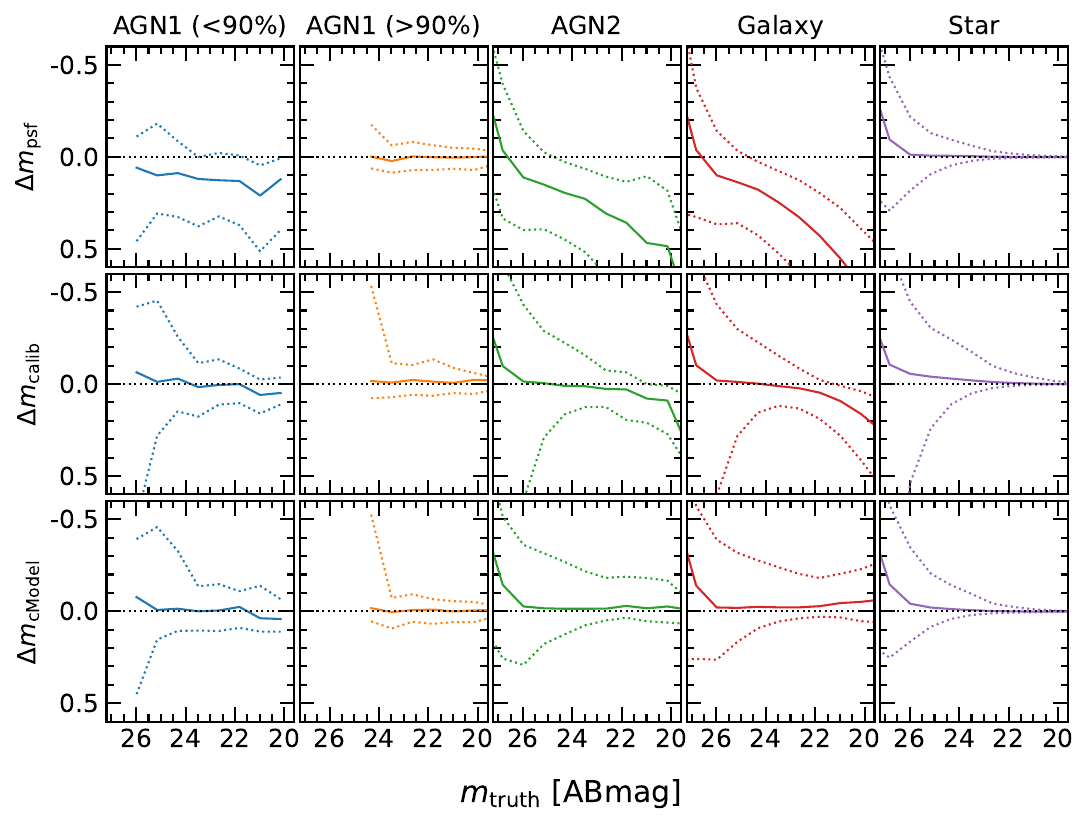}
    \caption{%
      Accuracy of the \ac{LSST} Science Pipelines flux estimators for different
      classes of objects. Each panel shows the median (10th and 90th
      percentile) magnitude difference between the measured flux and the truth
      flux in the $r$-band. Rows correspond to the flux estimated using
      \texttt{psfFlux}, \texttt{calibFlux} (defined as the 12 pixel aperture
      flux), and \texttt{cModelFlux}, respectively. Columns correspond to the
      truth labels, as shown in the top section.
    }
    \label{fig:flux_sigma_eta}
\end{figure*}

We evaluated the photometric accuracy across different classes of objects and
different flux definitions, notably the \texttt{psfFlux}, the
\texttt{calibFlux} (12 pixel aperture flux), and the \texttt{cModelFlux}.
We matched the objects with their truth catalog counterparts by minimizing the
combined difference in both measured position and flux. To quantify the
accuracy of the $r$-band flux $F_r$, we calculated both the \ac{NMAD}
$\sigma_\mathrm{NMAD} \equiv 1.48 \times |F_r - F_{r, \mathrm{truth}}| \,/\,
F_\mathrm{r, \mathrm{truth}}$ as well as the catastrophic outlier fraction
$\eta$ (\ie fraction of objects with $|F_r - F_{r,\mathrm{truth}}| \,/\,
F_{r,\mathrm{truth}} > 0.15)$.
Given that the measured flux uncertainties naturally increase towards fainter
fluxes, we calculated these statistics in three separate bins from
$r_\mathrm{truth}=20$ to $r_\mathrm{truth}=26$.
We summarize all our results in \cref{fig:flux_sigma_eta} and
\cref{tab:flux_sigma_eta}. Here we focus on the $20 < r < 22$ results in order
to highlight the systematic differences between the populations.

We start by discussing the stellar and galaxy populations. As expected, we
found stars best described by their measured \texttt{psfFlux}, with the lowest
$\sigma_\mathrm{NMAD} = 0.003$, across the different flux definitions. In
addition, we found $\eta = 0.001$ for all flux estimates.
On the contrary, we found galaxies best described by their \texttt{cModelFlux}
at $\sigma_\mathrm{NMAD} = 0.07$ (\texttt{cModelFlux}) compared to \eg
$\sigma_\mathrm{NMAD} = 0.108$ (\texttt{calibFlux}). Especially for bright
galaxies, the aperture fluxes tend to bias low due to the unaccounted for
extendedness of the source. Indeed, we observed the two estimates to agree
better towards fainter fluxes (see \cref{fig:flux_sigma_eta}).

\typei \acp{AGN} and their hosts form complicated morphologies from
pointlike-\acp{QSO} to galaxy-dominated systems with centrally concentrated
\ac{AGN} emission. To account for this, we measured separately the accuracy
for systems that are \ac{AGN}-dominated (host-contaminated), based on whether
the \ac{AGN} flux accounts for $>90\%$ ($<90\%$) of the total flux.
We found that apart from \texttt{psfFlux} for host-contaminated systems, all
flux estimates recovered the true flux on average without the biases present as
for bright galaxies. The effect of the host-galaxy contamination was
clearly detected in the measured \ac{AGN} \texttt{psfFlux} (top-left
panel of \cref{fig:flux_sigma_eta}), and in the high outlier fraction ($\eta =
0.477$, see \cref{tab:flux_sigma_eta}). We found the results for \typeii
\acp{AGN}, similar to the galaxy ones as is expected from \ac{AGN} obscuration.

Finally, in \cref{tab:flux_sigma_eta} we report on the flux accuracy on sources
split by morphology, defined as the extendedness as measured in the $r$-band.
Overall our findings remain unchanged. That is, pointlike sources were best
described by either their \texttt{psfFlux} ($\sigma_\mathrm{NMAD} = 0.071$) or
\texttt{cModelFlux} ($\sigma_\mathrm{NMAD} = 0.108$). For extended sources, we
found \texttt{cModelFlux} to perform the best ($\sigma_\mathrm{NMAD} = 0.071$),
followed by \texttt{calibFlux} ($\sigma_\mathrm{NMAD} = 0.108$). As expected,
the use of \texttt{psfFlux} resulted in highly biased estimates with $\eta =
0.546$.

These results highlight that at level of the flux definition, no single flux
definition captures fully and accurately the diverse \ac{AGN} population. Our
results indicate that overall the \texttt{cModelFlux} has the least amount of
problematic cases, while \texttt{psfFlux} is clearly preferred for point-like
sources. For \ac{AGN} population studies, a strategy utilizing a combination of
fluxes based on measured extendedness is preferred.

\subsection{Forced photometry, limiting magnitude, and extracted light curves}

We explored the accuracy of forced photometry catalog, based on forced
photometry as performed on the positions of the objects detected in the deep
coadded images. The forced photometry extraction was performed on each
single-visit image individually. We measured the photometric accuracy using
\texttt{psfFlux} and \texttt{psfFluxErr} from the forced photometry catalog,
and limiting the analysis to pointlike sources. We estimated the $5\,\sigma$
depth ($m_5$) as the magnitude at which $\SNR = 5$. For the $ugrizy$ bands
using forced photometry, we found
$m_{5,u} = {23.01}_{-0.39}^{+0.27}$,
$m_{5,g} = {24.00}_{-0.59}^{+0.33}$,
$m_{5,r} = {23.65}_{-0.45}^{+0.34}$,
$m_{5,i} = {23.28}_{-0.44}^{+0.33}$,
$m_{5,z} = {22.72}_{-0.38}^{+0.30}$,
$m_{5,y} = {21.68}_{-0.37}^{+0.32}$
(errors correspond to the 16th and 84th percentiles).

This is in overall good agreement with the \ac{LSST} error estimates as
forecast in \citet{ivezic2019ApJ...873..111I}. They provided an analytic
formula to compute $m_5$, which depends on the band, air mass, seeing, and sky
background \citep[][Eq.~6]{ivezic2019ApJ...873..111I}. To facilitate for an
accurate comparison between their work, we re-estimated their $m_5$ with the
median air mass, seeing, and sky background using the \ac{LSST} baseline v4.0
in COSMOS\@. We found $m_5 = (23.32, 24.25, 23.89, 23.41, 22.83, 21.93)$
($ugrizy$), which is within the errors of
\citet[][]{ivezic2019ApJ...873..111I}, but systematically brighter by
$0.1$--$0.3$ mag. Although part of this discrepancy can be attributed to
Galactic extinction, this effect is expected to be very small, if not
negligible, in a field like COSMOS\@. Moreover, the relatively large errors --
mostly attributed to variations in the air mass -- highlight the need to
exercise care when condensing the limiting magnitudes to a single number.

\begin{figure*}
    \includegraphics[width=\linewidth]{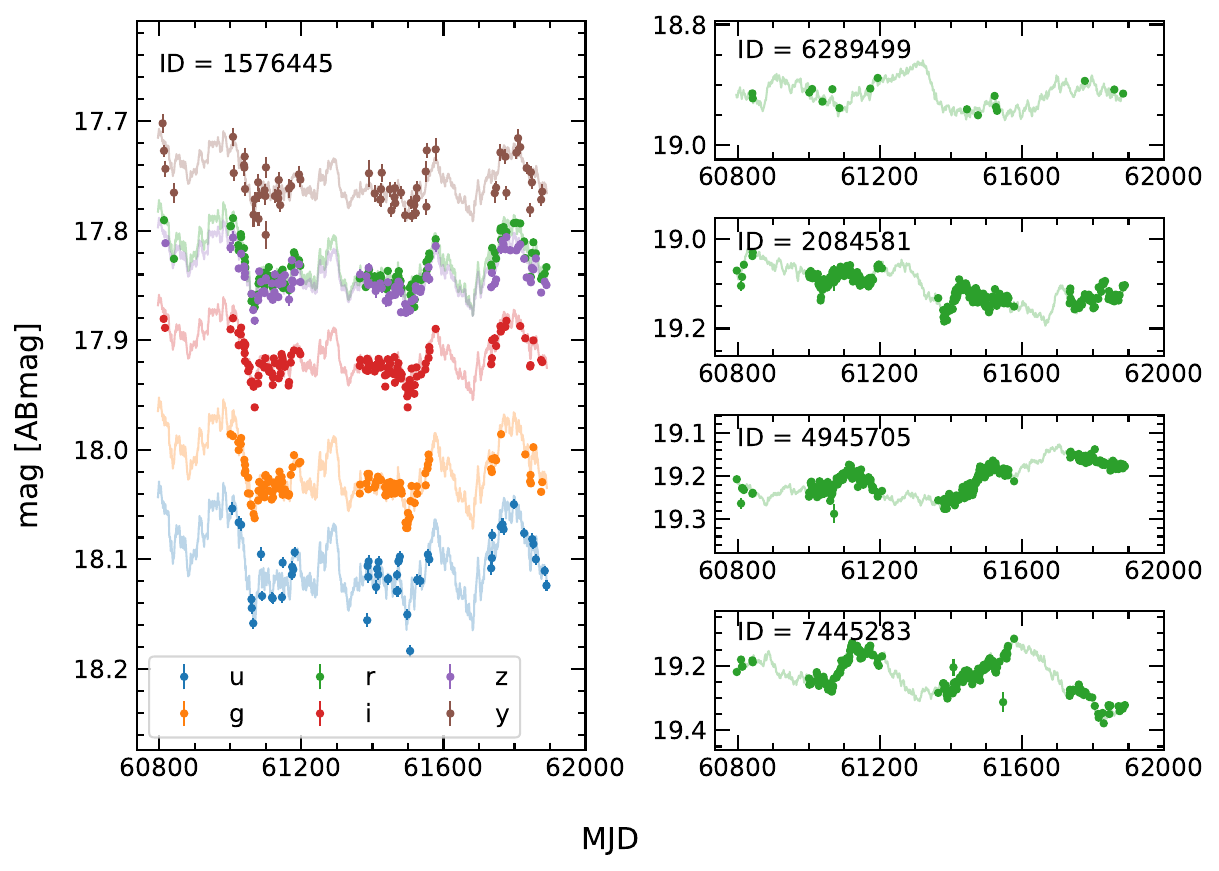}
    \caption{%
      Example light curves in the \ac{LSST} bands of bright \acp{AGN} within
      \ac{AGILE} DR1. The markers show \texttt{psfFlux} \ac{AGN} light curves
      from the forced photometry catalogs, while the lines correspond to the
      truth catalog light curves in the same bands as indicated by the colors.
      As explained in the text and \cref{app:dr1}, only the first exposure of
      each \ac{DDF} sequence is shown, and the measured light curves span the
      first three years of simulated \ac{LSST} operations in the COSMOS
      \ac{DDF} according to baseline v4.0. The right panels show light curves
      only in the $r$-band of four bright \acp{AGN}, while otherwise the axes
      and the units are the same as in the left panel. The top-right panel
      shows an \ac{AGN} ($\mathrm{ID} = 6289499$) that is located close  to the
      edge ($<1\arcsec$) of the detector layout (\cref{fig:detector_layout})
      and is sampled less frequently compared to the rest.
    }
    \label{fig:example_lightcurves}
\end{figure*}

As an example of the photometric accuracy using forced photometry, we
show the $ugrizy$ light curve of the brightest \ac{QSO} in the simulation in
\cref{fig:example_lightcurves}. Moreover, in the same figure, we also show the
$r$-band light curves of four bright \acp{AGN} from the simulation.

\section{AGN science applications}
\label{sec:science}

In this section, we move beyond using the photometric catalog solely for
testing technical aspects of the \ac{LSST} pipeline and explore its potential
for scientific applications in the context of \ac{AGN} science. In particular,
here we present the results we obtained by applying classical \typei \ac{AGN}
selection techniques based on color and variability.

In real \ac{LSST} observations, source classification will leverage the full
range of available data, likely through \ac{ML} approaches
\citep[\eg][]{Savic:2023}. However, this is not the goal of this
section, and the development of advanced techniques for identifying
\acp{AGN} using the \ac{AGILE} catalogs is deferred to future work. This
analysis provides an opportunity to further validate the catalog by applying
well-established \ac{AGN} selection techniques that are effective at
identifying bright, \ac{QSO}-like \typei \acp{AGN}. Moreover, it allows for the
quantification of the completeness and purity of these methods using an
intrinsically complete and well-characterized mock sample.

\subsection{AGN color-color selection}

\begin{figure*}
    \centering
    \includegraphics[width=\linewidth]{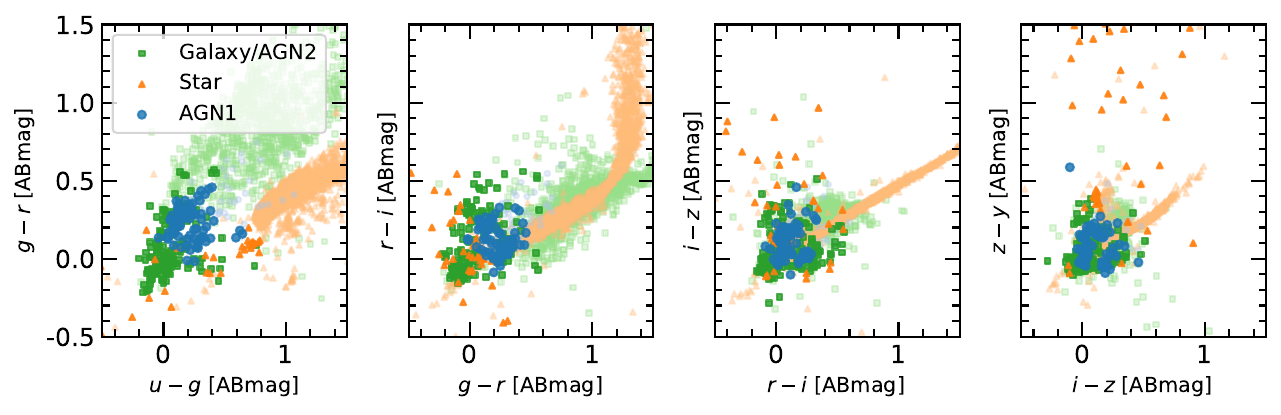}
    \caption{%
      Observed color-color diagrams using the \ac{LSST} bands for detected
      \ac{AGILE} sources. The markers and colors correspond labels of galaxies
      (or \typeii \acp{AGN}), stars, and \typei \acp{AGN} in accordance with
      the legend, and corresponding to the true labels. A darker color is used
      to mark sources selected using the \citet{Croom:2009} criteria. Only
      bright sources $g < 21.85$ and $i < 22$ are shown.
    }
    \label{fig:color_color_diagram}
\end{figure*}

Color-color diagrams have been widely used to identify \typei \acp{AGN} since
their discovery
\citep[\eg][]{%
  schmidt1983ApJ...269..352S,%
  Richards:2002,%
  Croom:2009,%
  euclid_matamoro2025arXiv250315320E%
}
as these sources occupy a distinct locus in the multi-dimensional color space,
separating them from inactive galaxies and main-sequence stars. Early selection
methods relied on morphology and simple $U-B$ bands cuts to identify sources
with an \ac{UV} excess, due to the blue bump of the accretion disk
\citep[\eg][]{Sandage:1965, schmidt63}. Over time, the increasing availability
of multi-band photometry and a refined understanding of the \ac{AGN} locus have
led to more sophisticated color-based selection criteria
\citep[\eg][]{Fan:1999, Richards:2002, Croom:2009}, significantly improving the
efficiency and completeness of \ac{AGN} identification.

Towards this end, we first selected objects from the photometric
catalog based on the three-year coadded images. For these objects, we
then applied the selection criteria proposed by
\citet[][Eqs.~1--3]{Croom:2009} to identify \acp{QSO} in the 2SLAQ survey at
$z<2.6$. Since the criteria by \citet{Croom:2009} are designed to select
luminous \acp{QSO}, they include relatively strict cuts in observed flux,
limiting the selection to bright sources with $g<21.85$ and $i<22$. These
constraints significantly narrow the \ac{AGILE} DR1 sample: $120$
\typei \acp{AGN} (out of $685$ detected in the object table), $37$ \typeii
\acp{AGN} (out of $9487$), $1186$ galaxies (out of $253\,243$), and $3184$
stars (out of $27\,345$) satisfy the selection. In
\cref{fig:color_color_diagram}, we show the color-color diagrams for these
sources. Following \citet{Croom:2009}, we use \ac{PSF} magnitudes.

Among the $120$ bright \typei \acp{AGN}, $116$ lie at $z < 2.6$, of which $77$
are correctly identified using the \citet{Croom:2009} selection criteria.
However, the same criteria incorrectly classify $48$ stars, $140$ galaxies, and
eight \typeii \acp{AGN} as \typei. Restricting the sample to sources with $g <
21.85$ and $i < 22$, the \citet{Croom:2009} criteria yield a completeness of
$65\%$ ($67\%$ when considering only $z < 2.6$) but a high contamination rate
of $72\%$.

To compare with \citet{Croom:2009}, we split our sample into two $g$-band
magnitude bins. For $g \leq 20.5$, we recover $30$ out of $31$ \acp{AGN},
consistent with the ${\sim} 100\%$ recovery reported for very bright sources,
although the contamination remains high at $63\%$. For $20.5 < g < 22$, we
correctly identify $47\,/\,84$ \typei \acp{AGN}, corresponding to a slightly
lower completeness than the ${\sim} 70\%$ reported by \citet{Croom:2009} for
the same range, with a contamination of $75\%$.

Following \citet{Croom:2009}, we did not apply any morphological selection.
However, introducing the additional requirement for the sources to be
point like in their reference band ($\texttt{refExtendedness} = 0$), although
reducing the number of \typei \acp{AGN} which satisfy all conditions down to
$57$ ($45$ of which are correctly selected), effectively removes all galaxy and
\typeii \ac{AGN} contaminants, dropping the contamination to $43\%$.

\subsection{Variability analysis}

The \ac{AGN} optical continuum variability is a key feature for their discovery
as it has proven to be an effective method for distinguishing \typei \acp{AGN}
from other types of sources
\citep[\eg][]{%
  van-den-Bergh:1973,%
  Koo:1986,%
  Sanchez-Saez:2019,%
  De-Cicco:2021,%
  Savic:2023,%
  paolillo2025NCimR..48..537P%
}.
Here we present a simple test to quantify the \ac{LSST} \ac{AGN} recovery
performance based solely on the variability of the sources. Nowadays, \ac{AGN}
selection via variability relies on computing a large set of features from the
light curves \citep[\eg] []{Savic:2023,De-Cicco:2025}; however, for simplicity,
we adopt a single low-statistics variability metric, the \ac{RMS} deviation
\citep[\eg][]{Sesar:2007,Trevese:2008a,Sarajedini:2011,Pouliasis:2019}.

Since for extended sources, \ac{PSF} fluxes exhibit artificial variability due
to variations in the \ac{PSF}, atmospheric conditions and positional
inaccuracies, for this analysis we use \texttt{psfFlux + psfDiffFlux}, where
\texttt{psfFlux} is the (constant) flux from the coadded template image, and
\texttt{psfDiffFlux} is the difference flux compared to the template image. For
\acp{AGN}, this accounts for both the seeing, and the host galaxy (depending on
the deblending quality),
while for galaxies and stars this expression reduces on average to
\texttt{psfFlux}. We restrict the analysis to sources detected in the $r$-band
in at least ten separate epochs, with detections defined by a signal-to-noise
ratio of $\SNR > 5$. We also required sources to have no failures in their
\texttt{psfFlux} and extendedness measurements (\ie $\texttt{r\_psfFlux\_flag}
= \texttt{r\_extendedness\_flag} = 0$ in the object table). Finally, to
minimize photometric issues, we excluded sources located near the survey edges
and in the proximity of bright, saturated, stars \citep[\eg][]{Poulain:2020}.
Specifically, we excluded an annular region within $2\arcmin$ of the survey
boundary and masked objects within $2\arcmin$ of stars with $r < 9$ and within
$1\arcmin$ of stars with $r < 11.5$. These final cuts reduced the final
analyzed sample by ${\sim} 10\%$.

The \ac{RMS} can be strongly affected by photometric outliers. To
account for this, we clipped the light curves by excluding points more than
$5\,\sigma$ away from the median magnitude.
We investigated this outlier fraction in terms of relevant simulation
parameters. We found that a single exposure has an elevated outlier fraction
mainly due to a low number of \ac{PSF} stars ($N_\mathrm{PSF}$). We used $r <
24$, point-like, and non-flagged detections with $N > 30$ observations to
quantify the outlier fraction. We found that a typical exposure
($N_\mathrm{PSF} = 124$) has an outlier fraction of $0.4\%$. At a low
percentile ($N_\mathrm{PSF} < 40$ or $1\%$), this fraction elevates
steeply to $10\%$ and above. Moreover, we find that $N_\mathrm{PSF}$ and
seeing $\theta$ are strongly anti-correlated ($R = -0.95$). We find $\theta <
2\arcsec$ to yield securely $N_\mathrm{PSF} > 40$, while $N_\mathrm{PSF} =
124$ corresponds to $\theta \approx 1\arcsecf3$. These findings have
important implications for \ac{LSST} observations, and care should be taken
in using seemingly good-quality (\ie non-flagged) observations for
variability analyses.

\begin{figure}[htbp]
    \centering
    \includegraphics[width=\linewidth]{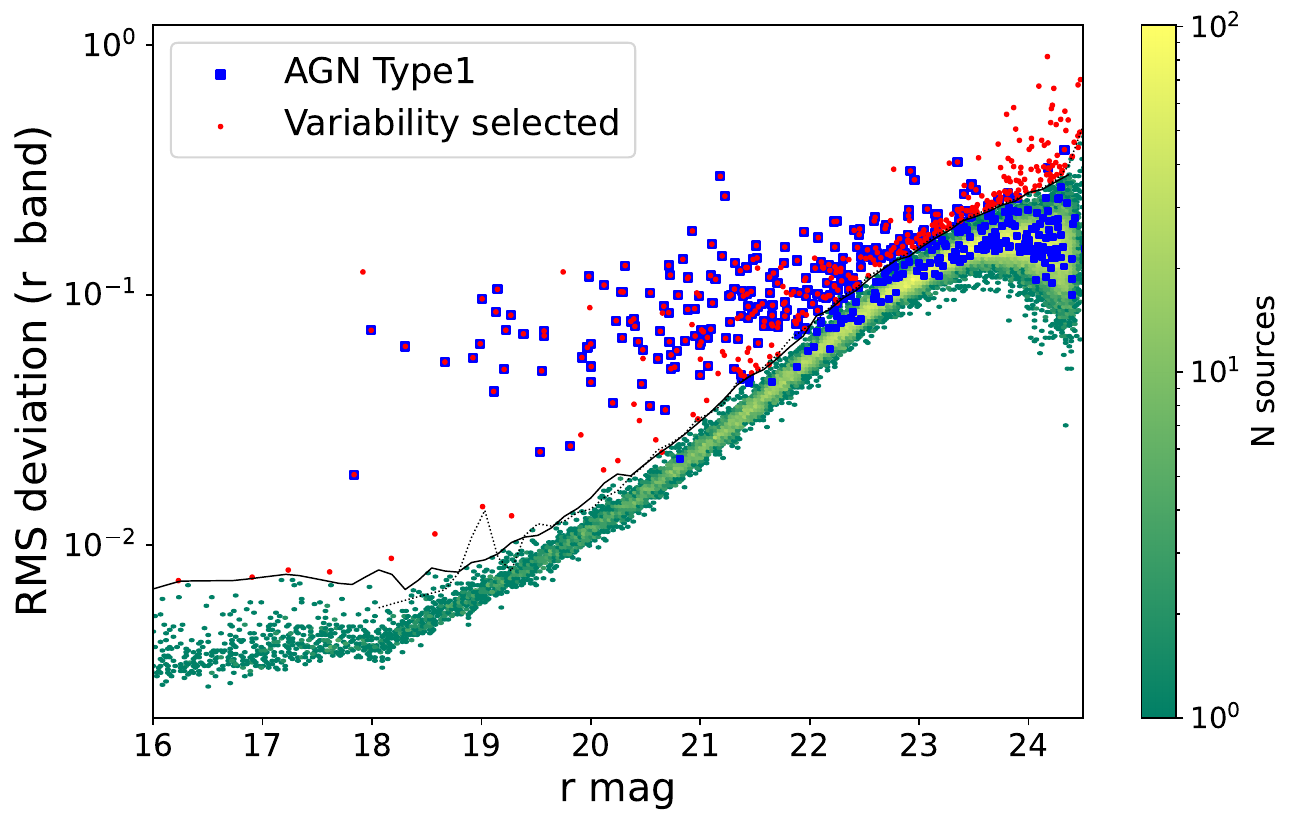}
    \caption{%
      Magnitude vs. \ac{RMS} deviation in the $r$-band. Sources which are more
      than three times the \ac{MAD} above the median of the distribution in a
      magnitude bin are selected as variable and labeled with red markers. Blue
      squares indicate \typei \acp{AGN}. The solid and dotted black lines
      represent the thresholds used to select variable sources for point-like
      and extended sources respectively.
    }
    \label{fig:rms_plot}
\end{figure}

In \cref{fig:rms_plot} we show the \ac{RMS} as a function of  \texttt{psfFlux}
magnitude. A clear trend is visible: fainter sources exhibit higher \ac{RMS}
values due to the increasing relative contribution of background noise, while
at bright ($r < 18$) magnitudes the \ac{RMS} flattens at ${\sim}
0.005\,\mathrm{mag}$, consistent with the typical level of systematic
uncertainty. Assuming that the vast majority of sources are non-variable, we
computed the median observational \ac{RMS} and its \ac{MAD}\footnote{%
  Here we define the \ac{MAD} following \citet[][Eq.~1]{Pouliasis:2019}, \ie
  with the $1.4826$ factor already included to scale it to the standard
  deviation.
}
in bins of magnitude. Following \citet{Pouliasis:2019}, we selected as variable
those sources with an \ac{RMS} exceeding the median by more than three times
their \ac{MAD}. We adopted the lower threshold of three (rather than three and
a half) because the much higher number of epochs in our dataset makes the
\ac{RMS} converge more reliably to the true, magnitude dependent, value
\citep[\eg][]{Trevese:2008a}. We performed this method separately for extended
and point-like sources \citep[\eg][]{Sarajedini:2011, Pouliasis:2019}. All
sources with $r < 16$ were treated as non-variable since their variability
measurements are affected by saturation.

The recall power of this selection method depends strongly on source magnitude.
Overall, we recover $53\%$ of \typei \acp{AGN} present in the catalog after
applying the quality cuts ($225\,/\,421$). However, the recovery fraction
varies significantly with brightness: $98\%$ ($64\,/\,65$) for sources with $r
\leq 21$, $91\%$ ($68\,/\,75$) for $21 < r \leq 22$, and only $33\%$
($93\,/\,281$) for $r > 22$, where the typical \ac{AGN} variability amplitudes
become comparable to the observational scatter. Compared to
\citet{Trevese:2008a}, who reported a completeness of $44\%$ for $V < 24$
sources, based on an eight-epoch survey with the Wide Field Imager,  we find a
higher recall. Restricting our analysis to the same magnitude range (though in
a different band), we obtain an overall completeness of $61\%$ ($223\,/\,366$).
This improvement is not surprising given the higher signal-to-noise ratio at
the adopted magnitude limit and the higher LSST sampling rate.

We also investigated the dependence of completeness on physical parameters,
finding that the selection preferentially identifies systems hosting more
massive black holes and/or accreting at higher Eddington ratios. This trend
primarily reflects the imposed magnitude limit, rather than the variability
prescription described in Sec. \ref{sec:agnvariability}. Within a fixed
redshift bin, a detectability cut such as $r < 22$ effectively corresponds to a
luminosity threshold, which translates into a bias toward higher black hole
masses and/or higher Eddington ratios. Although the adopted variability
prescription assigns larger intrinsic amplitudes to lower BH mass and less
luminous AGN, this effect is subdominant compared to the bias introduced by the
magnitude limit.

An analogous trend with magnitude is found in the contamination fraction:
$27\%$, $31\%$, and $63\%$ in the same magnitude bins, with an overall
contamination of $48\%$. When restricting to $r < 24$, we find a contamination
of $42\%$, which is slightly higher than the $<40\%$ reported by
\citet{Trevese:2008a}. \citet{Pouliasis:2019} also found magnitude-dependent
contamination which is consistent with our trend: as low as $30\%$ for
$z_\mathrm{HST} < 23$ and $>70\%$ at fainter magnitudes. Unlike completeness,
contamination metrics behave very differently for extended and point-like
sources. For extended sources, contamination is already $40\%$ at $r < 22.5$
and quickly rises to $>90\%$ at fainter magnitudes. For point-like sources, the
average contamination is $10\%$ at $r < 23.5$ (or $6\%$ if bright stars with $r
< 18$ classified as variable, see \cref{fig:rms_plot}, are excluded). However,
at fainter magnitudes, the contamination also steeply approaches $90\%$.

Finally, we note that this selection method proved completely ineffective at
recovering \typeii \acp{AGN}, with a completeness $<0.3\%$ which is comparable
with the recover rate of non variable sources in the catalog. This is
unsurprising from single-band data alone since \typeii \acp{AGN} have
suppressed optical variability, and it is therefore much more challenging to
recover them. Instead, variability selection including multiple bands, colors,
and morphology should be explored to improve on the selection of \typeii
\acp{AGN}.

\section{Summary and conclusions}
\label{sec:conclusions}

We presented \ac{AGILE}, an end-to-end simulation pipeline of \acp{AGN},
galaxies, and stars, designed to forward model accurately these populations in
the \ac{LSST} survey. \ac{AGILE} first builds a mock catalog including
\acp{AGN}, galaxies, and stars based on empirical relations, ensuring
consistency with observed \ac{AGN} and galaxy properties. \ac{AGILE} then
includes a model for the instrumental effects, survey design, and time-domain
variability for all sources. This enables a dynamic, evolving representation of
the \ac{AGN} population as it will be observed by \ac{LSST}, providing an
essential framework for optimizing \ac{AGN} detection and classification.

We also release the \ac{AGILE} \ac{DR1} which consists of a
$24\,\mathrm{deg}^2$ underlying mock truth catalog, and $1\,\mathrm{deg}^2$ of
simulated images and photometric catalogs corresponding to a total of
$\nvisitdri$ visits over three years (\cref{app:dr1}). Specifically, this
dataset resembles the first \ac{LSST}
\ac{DP1}\footnote{\url{https://dp1.lsst.io/}} in the \ac{ECDFS} taken with the
\ac{LSST} Commissioning Camera ($\ac{FOV} \approx 0.5\,\mathrm{deg}^2$). The
\ac{LSST} \ac{DP1} \ac{ECDFS} survey depth is comparable to \ac{AGILE} \ac{DR1}
with $855$ visits compared to $\nvisitdri$, but the baseline is considerably
shorter at roughly a month and a half compared to three years. Thus \ac{AGILE}
is specifically suited for early scientific exploitation of these \ac{LSST}
data.

By exploring first these \ac{AGILE} data, we conclude that:
\begin{itemize}

    \item Based on the literature results, we find that our truth population of
      \typei and \typeii \acp{AGN}
      is consistent with both the \ac{XLF}, as well as the
      optical $g$-band number counts, and the quasar luminosity function in the
      $B$-band across a wide baseline in redshift $z < 4$. The mock \ac{AGN}
      population also reproduces the local \ac{BHMF}.

    \item In the LSST Science Pipelines, we find that no single flux estimator
      is fully able to capture accurately the input flux of the diverse
      \ac{AGN} population mostly attributed to the varying \ac{AGN}
      contribution and host-galaxy morphology. Therefore, separate flux
      estimators should be utilized, for example for \ac{AGN} and host-galaxy
      systems classified as point like or extended. Specifically, our results
      suggest utilizing \texttt{psfFlux} and \texttt{cModelFlux} for point-like
      and extended sources, respectively.

    \item We applied the standard \typei \ac{AGN} color-color selection by
      \citep{Croom:2009} for bright sources (\ie $g < 21.85$ and $i < 22$)
      finding an overall completeness of $67\%$. In agreement with the results
      by \citep{Croom:2009}, the completeness is as high as ${\sim} 100\%$ for
      very bright sources ($g \leq 20.5$) and  ${\sim} 55\%$ for fainter
      sources ($20.5 < g < 22$). However, contamination is as high as ${\sim}
      70\%$ in both magnitude bins and drops to ${\sim} 40\%$ only when
      restricting the analysis to point-like sources.

    \item We also applied a simple \typei \ac{AGN} variability selection based on
      the \ac{RMS} deviation finding that the recovery fraction varies
      significantly with brightness: $98\%$ for sources with $r \leq 21$,
      $91\%$ for $21 < r \leq 22$, and only $33\%$ for $r>22$, where the
      typical \ac{AGN} variability amplitudes become comparable to the
      observational scatter expected from a single-visit \ac{LSST} exposure. We
      also find that contamination is as low as $\lesssim 10\%$ for point-like
      sources at $r < 23.5$  ($<40\%$ for extended sources at $r<22.5$), while
      rising to $90\%$ at $r > 23.5$.

\end{itemize}

\noindent
The full \ac{AGILE} dataset discussed here provides lucrative opportunities for
further analysis. It is an ideal test-bench for studying \eg \ac{AGN}
photometric redshifts, \ac{AGN} host-galaxy morphologies, and the physical
parameter estimation of \ac{AGN} and their host galaxies in \ac{LSST}, \Euclid,
and beyond.

\section*{Data availability}

The main \ac{AGILE} portal is \url{https://www.oa-roma.inaf.it/lsst-agn/}. The
\ac{AGILE} source code is hosted in the INAF GitLab
\url{https://www.ict.inaf.it/gitlab/akke.viitanen/lsst_agile}. Access to the
AGILE DR1 repository is granted upon a reasonable request to the authors.

\begin{acknowledgements}

We thank the \ac{INAF} computing system PLEIADI, for the availability of
high-performance computing resources and support.
We acknowledge the use of the ADHOC (Astrophysical Data HPC Operating Center)
resources, within the project ``Strengthening the Italian Leadership in ELT and
SKA (STILES)'' proposal nr. IR0000034, admitted and eligible for funding from
the funds referred to in the D.D.~prot.~no.\ 245 of August 10, 2022 and D.D.\
326 of August 30, 2022, funded under the program ``Next Generation EU'' of the
European Union, ``Piano Nazionale di Ripresa e Resilienza'' (PNRR) of the
Italian Ministry of University and Research (MUR), ``Fund for the creation of
an integrated system of research and innovation infrastructures'', Action 3.1.1
``Creation of new IR or strengthening of existing IR involved in the Horizon
Europe Scientific Excellence objectives and the establishment of networks''
A.V.\ acknowledges support from the Finnish Academy of Science and Letters and
the Foundations' Post Doc Pool.
A.B.\ acknowledges the hospitality of the University of Geneva.
M.P.\ acknowledges support from the Italian PRIN - MIUR 2022 ``SUNRISE'' and
the INAF grant TIMEDOMES\@. The research leading to these results have received
funding by the EU HORIZON-MSCA-2023-DN Project 101168906 `TALES: Time-domain
Analysis to study the Life-cycle and Evolution of Supermassive black holes`
DD acknowledges PON R\&I 2021, CUP E65F21002880003, and Fondi di Ricerca di
Ateneo (FRA), linea C, progetto TORNADO.
D.I.\ and A.B.K.\ acknowledge funding provided by the University of Belgrade --
Faculty of Mathematics through the grant (the contract
451-03-136/2025-03/200104) of the Ministry of Science, Technological
Development and Innovation of the Republic of Serbia.
G.D.S.\ acknowledges support from Gaia DPAC through INAF and ASI (PI: M. G.
Lattanzi), and from INFN (Naples Section) through the QGSKY and Moonlight2
initiatives.
RJA was supported by FONDECYT grant number 1231718 and by the ANID BASAL
project FB210003.
CR acknowledges support from SNSF Consolidator grant F01$-$13252, Fondecyt
Regular grant 1230345, ANID BASAL project FB210003 and the China-Chile joint
research fund.
GP and MT acknowledge funding by the European Union -- NextGenerationEU and by
the University of Padua under the 2023 STARS Grants@Unipd programme
(``CONVERGENCE'' project).

\end{acknowledgements}

\bibliography{aa58627-25}
\bibliographystyle{aa}

\begin{appendix}

\section{AGILE pipeline flowchart}
\label{sec:agile_pipeline_flowchart}

\begin{sidewaysfigure*}
  \centering
  \includegraphics[width=1.0\textwidth]{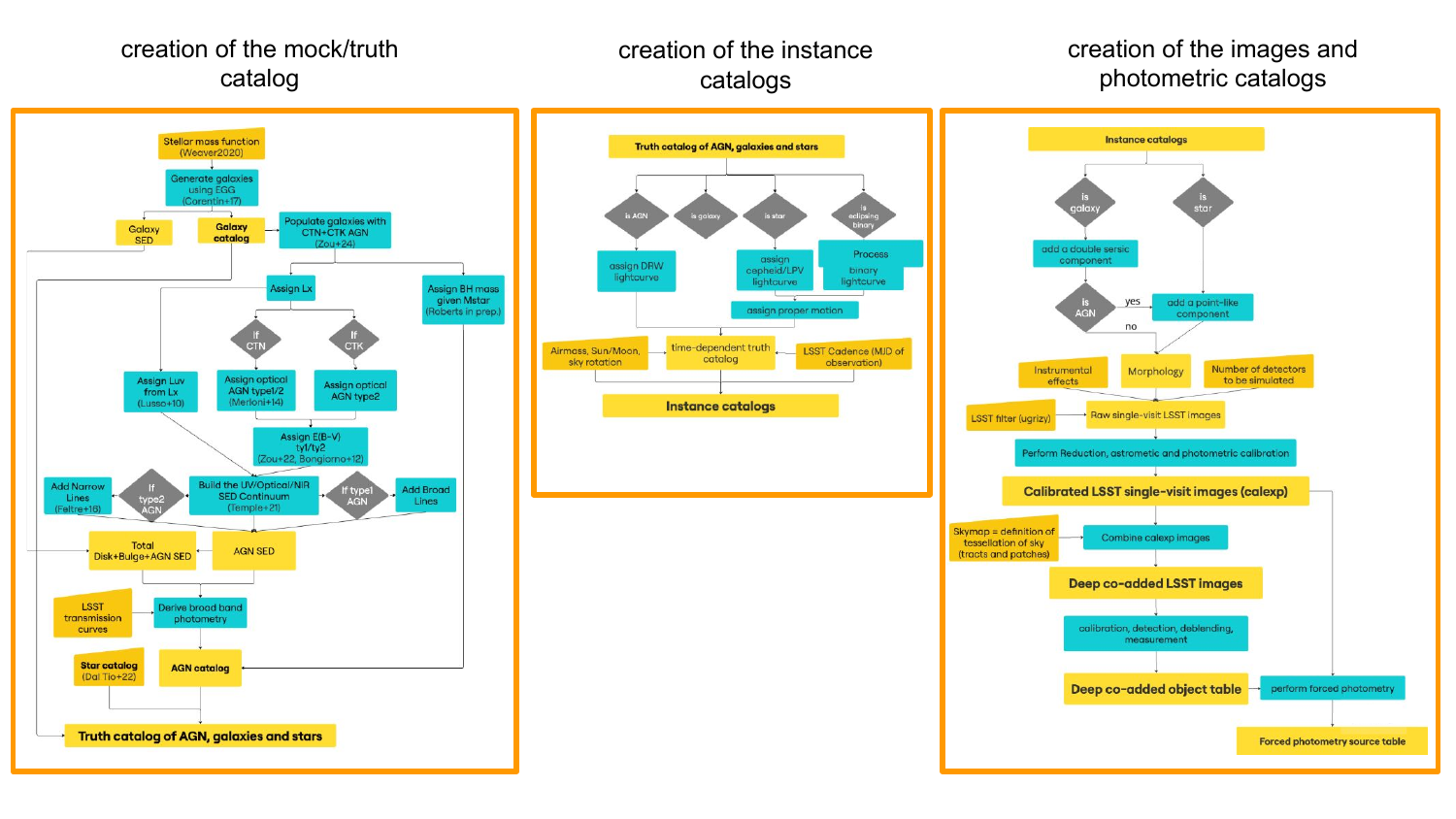}
  \caption{%
    The AGILE pipeline flowchart. Blocks from left to right describe the three
    main steps of the simulation: 1) The creation of the mock truth catalog
    which includes \acp{AGN}, galaxies and stars; 2) the creation of the
    instance catalogs obtained from the truth catalog by including variability
    information for \acp{AGN} and stars; 3) the creation of \ac{LSST}
    single-visit images according to the \ac{LSST} survey strategy,
    the reduction, and coaddition performed using the \ac{LSST} Science
    Pipelines. This step also includes the photometric analysis (including
    forced photometry) which leads to the final photometric catalogs.
  }
  \label{fig:agile}
\end{sidewaysfigure*}

We show the AGILE pipeline flowchart in \cref{fig:agile}.

\section{Accuracy of extrapolation in $p(\lambda_\mathrm{SAR})$}
\label{app:plambda_extrapolation}

\begin{figure*}[htbp]
  \centering
  \includegraphics[width=0.85\linewidth]{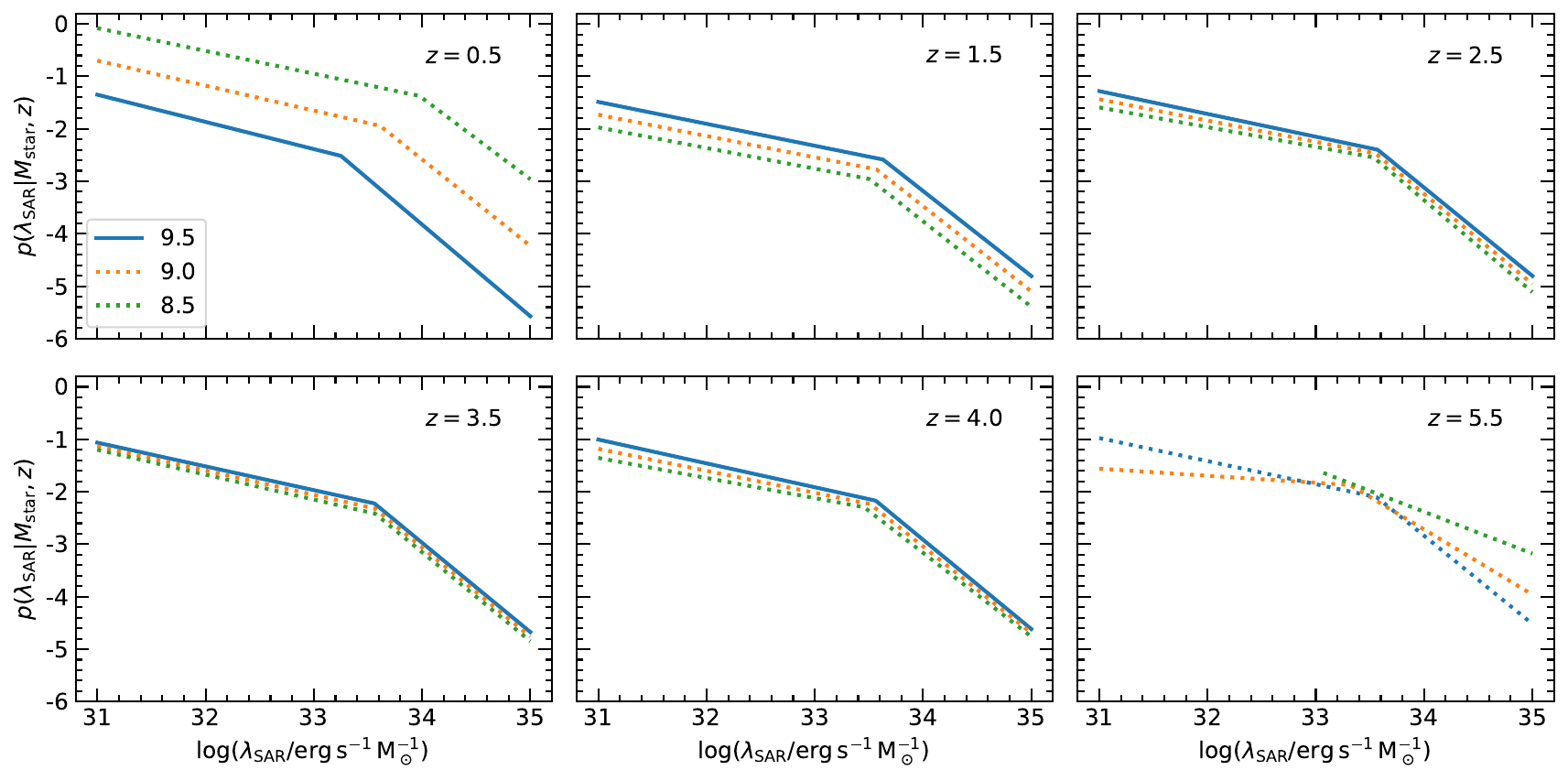}
  \caption{%
    Accuracy of extrapolation in the assumed \plambdafull \citepalias{Zou:2024}
    using the parameter maps. The panels show different $z$, while the lines
    correspond to different $M_\mathrm{star}$ as indicated by the legend.
    Dotted lines indicate extrapolation beyond the formal limits $\logMstar <
    9.5$ or $z > 4.0$ using \texttt{scipy.interpolate.RegularGridInterpolator}.
    Given the uncertainties of the extrapolation (indicated most clearly in the
    top-left panel), in this work we choose to extrapolate by using they
    boundary values $\logMstar = 9.5$ and/or $z=4.0$, corresponding to the blue
    solid curve (where applicable).
  }%
  \label{fig:test_zou_extrapolation_accuracy}
\end{figure*}

As explained in \cref{sec:accretion_rate_distribution}, in order to assign
$\lambda_\mathrm{SAR}$ to a complete $M_\mathrm{star} > 10^{8.5}\,M_\odot$
population of galaxies, we need to choose an extrapolation scheme. Choices
for the extrapolation include boundary extrapolation (choosing the
minimum and/or maximum value) or extrapolation with splines functions of
varying degrees. In this work, we choose to extrapolate the parameter maps of
\citetalias{Zou:2024} by imposing a minimum $M_\mathrm{star} =
10^{9.5}\,M_\odot$ and maximum $z = 4$ where applicable. We compare this
choice to the extrapolation scheme of
\texttt{scipy.interpolate.RegularGridInterpolator} \citep[][version
1.16.2]{scipy2020SciPy-NMeth}. This choice has the most consequences at $z =
0.5$, where differences in the chosen extrapolation can lead to a difference
of $2.0\,\mathrm{dex}$ in $p(\lambda)$, while the magnitude of the difference
is smaller at higher $z$ (see \cref{fig:test_zou_extrapolation_accuracy}).

\section{\texttt{QSOGEN} posterior distribution parameters}
\label{app:qsogen}

\begin{figure*}
    \centering
    \includegraphics[width=0.65\linewidth]{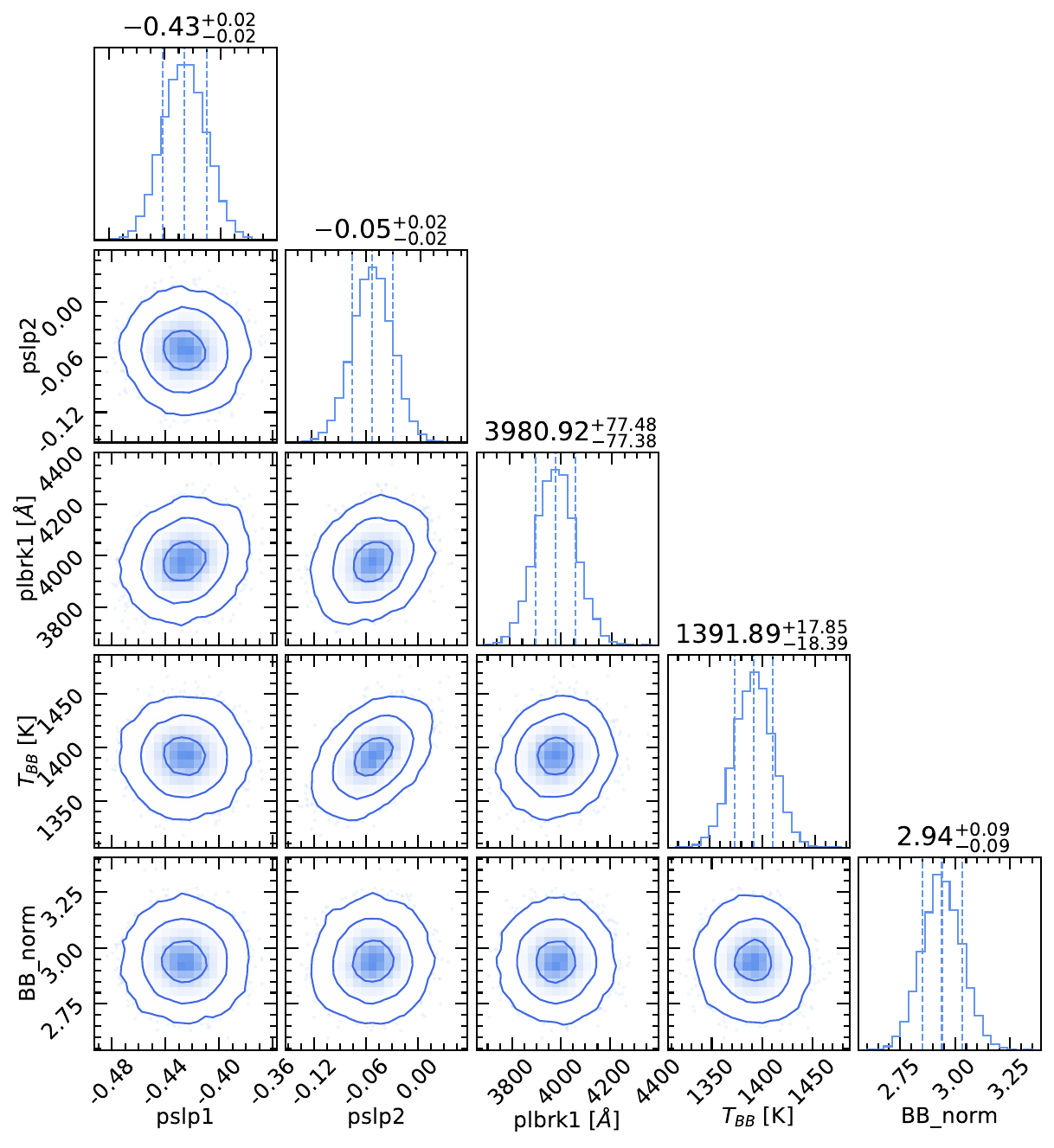}
    \caption{%
      The posterior distribution of the five free \textsc{QSOGEN} parameters
      (\cref{sec:qsogen}), derived from a pure quasar sample with $\logLopt
      \geq 45.5$. The contours represent the $1\,\sigma, 2\,\sigma$, and
      $3\,\sigma$ confidence levels. The vertical lines show the 16\%, 50\%,
      and 84\% percentiles, whose values are reported above each panel.
    }
    \label{fig:posterior}
\end{figure*}

We show the \texttt{QSOGEN} posterior distribution parameters
(\cref{sec:qsogen}) in \cref{fig:posterior}.

\section{AGILE first data release}
\label{app:dr1}

\begin{figure}[htbp]
  \centering
  \includegraphics[width=\linewidth]{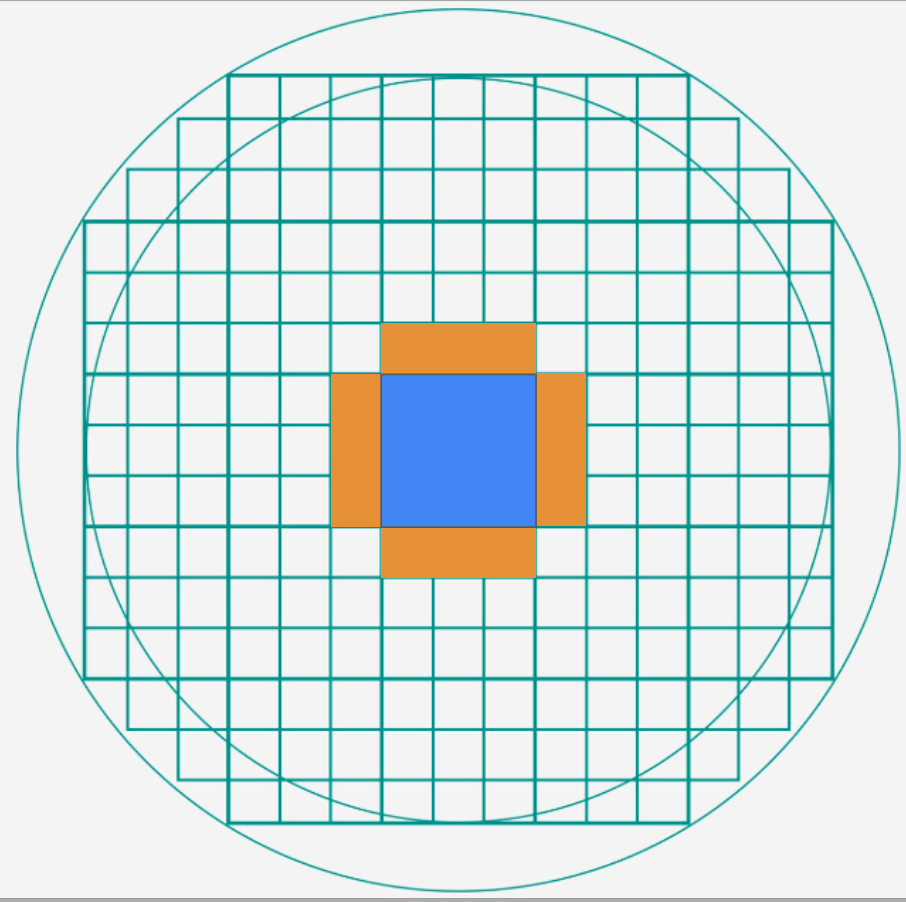}
  \caption{%
    The LSSTCam detector layout assumed in \ac{AGILE} \ac{DR1}. The LSSTCam
    consists of $189$ detectors covering the full $9.6\,\mathrm{deg}^2$
    \ac{FOV} (the grid), and the commissioning camera used for \ac{LSST}
    \ac{DP1} corresponds to nine detectors (blue region). For the purpose of
    reducing computational cost, for DR1 we choose to simulate a subset
    consisting of the central $21$ detectors corresponding to a single-visit
    area of $1.1\,\mathrm{deg}^2$ (blue and orange regions).
  }%
  \label{fig:detector_layout}
\end{figure}

Here we describe \ac{AGILE} DR1, which is an application of the software in
order to create an \ac{LSST} like array of simulated catalog-level products and
images. The simulation has been run under Comitato di Allocazione di Tempo di
calcolo e Spazio di archiviazione INAF (CAT\&S) Pleiadi Call 5 with a total of
$480\,000$ core hours awarded on the PLEIADI system of Trieste, Italy
\citep{taffoni2020ASPC..527..307T,bertocco2020ASPC..527..303B}. The \ac{AGILE}
DR1 consists of the following output products:
\begin{itemize}

  \item Truth catalog of \acp{AGN}, galaxies and stars with an area of
    $24\,\mathrm{deg}^2$, $0.2 < z < 5.5$, and $\logMstar > 8.50$, assuming the
    COSMOS2020 stellar mass function, and the specific accretion rate
    distribution from \citetalias{Zou:2024}
    (\cref{sec:galaxy,sec:agn,sec:star})

  \item Truth \acp{SED} of \acp{AGN} and galaxies in the optical/\ac{UV} and
    \ac{NIR} from $500\,\angstrom$ to $25\,\micron$ (\cref{sec:agn}),

  \item Truth \ac{AGN} $ugrizy$ light curves with ten-year baseline and
    resolution of one day (\cref{sec:variability}),

  \item Single-visit \texttt{calexp} and deep coadded $ugrizy$ simulated
    \ac{LSST} images corresponding approximately to $1\,\mathrm{deg}^2$ ($21\,/\,189$
    detectors, \cref{fig:detector_layout}), and the first three years of
    \ac{LSST} in COSMOS ($\nvisitdri$ visits) using the modified baseline 4.0
    (see \cref{sec:image}),\footnote{%
      In detail, in the original baseline 4.0 at each \ac{DDF} visit, a
      sequence of $N_{ugrizy} = (8, 10, 20, 20, 24, 18)$ exposures is typically
      taken back-to-back. Instead, in the modified baseline 4.0, we choose to
      only simulate the first visit of each such sequence. This optimization
      reduces the total computation time at the expense of the final coadded
      depth without compromising the cadence at which the \ac{AGN} light curves
      are sampled.
    }

  \item Object catalog extracted from the deep coadded images
    ($1\,\mathrm{deg}^2$, $\nvisitdri$ visits), including the best-match
    counterpart in the truth catalog (see \cref{sec:pipeline}),

  \item Forced photometry catalog extracted from the single-visit
    \texttt{calexp} images ($1\,\mathrm{deg}^2$, $\nvisitdri$ visits) at the
    positions of the object catalog detections (see \cref{sec:pipeline}),

  \item The best-matching counterpart for each object catalog entry in the
    truth catalog.

\end{itemize}
In addition, we provide python software (Jupyter notebooks) which demonstrates
the usage and analysis of these data products. Updates concerning the \ac{AGILE} DR1
and subsequent datasets are communicated through the main \ac{AGILE} portal
(\url{https://www.oa-roma.inaf.it/lsst-agn/}.

\section{Catalog column descriptions}
\label{app:columns}

We show a complete listing of the truth catalog parameters in
\cref{tab:catalog_columns}. The columns in the photometric catalogs are
detailed in the LSST Science Pipelines Software document
\citep{rubin202510.71929}, Rubin Data Product definition document
(\url{https://lse-163.lsst.io/}), and in the \ac{LSST} \acl{DP0.2}
documentation (\url{https://dp0-2.lsst.io/data-products-dp0-2/index.html}).

\begin{table*}[htbp]
\caption{List of parameters included in the truth catalog.}
\label{tab:catalog_columns}
\begin{tabular}{rlll}
  \hline\hline
ID & Name & Description & Unit \\
\hline
$  0$ & ID                              & Unique ID                                                                 & \ldots       \\
$  1$ & RA                              & Right ascension (J2000)                                                   & deg          \\
$  2$ & DEC                             & Declination (J2000)                                                       & deg          \\
$  3$ & Z                               & Cosmological redshift                                                     & \ldots       \\
$  4$ & D                               & Luminosity distance or distance modulus for stars                         & Mpc or mag   \\
$  5$ & M                               & $\logten$ of stellar mass                                                 & $M_\odot$    \\
$  6$ & SFR                             & Star-formation rate                                                       & $M_\odot\,\mathrm{yr}^{-1}$ \\
$  7$ & PASSIVE                         & Is passive (non-star-forming)                                             & \ldots       \\
$  8$ & CLUSTERED                       & Is clustered                                                              & \ldots       \\
$  9$ & DISK\_ANGLE                     & Galaxy disk position angle                                                & deg          \\
$ 10$ & DISK\_RADIUS                    & Galaxy disk half-light radius                                             & arcsec       \\
$ 11$ & DISK\_RATIO                     & Galaxy disk ratio of minor to major axis                                  & \ldots       \\
$ 12$ & BULGE\_ANGLE                    & Galaxy bulge position angle                                               & deg          \\
$ 13$ & BULGE\_RADIUS                   & Galaxy bulge half-light radius                                            & arcsec       \\
$ 14$ & BULGE\_RATIO                    & Galaxy bulge ratio of minor to major axis                                 & \ldots       \\
$ 15$ & AVLINES\_BULGE                  & Emission line extinction in the bulge                                     & mag          \\
$ 16$ & AVLINES\_DISK                   & Emission line extinction in the disk                                      & mag          \\
$ 17$--$40$ & band\_disk                & Galaxy disk band flux in the observer frame                               & $\mu$Jy      \\
$ 41$--$64$ & band\_bulge               & Galaxy bulge band flux in the observer frame                              & $\mu$Jy      \\
$ 65$--$88$ & magabs\_band              & Galaxy band absolute magnitude in the source's rest frame                 & ABmag        \\
$ 89$ & log\_lambda\_SAR                & $\logten$ of specific accretion rate                                      & $\mathrm{erg}\,\mathrm{s}^{-1}\,M_\odot^{-1}$ \\
$ 90$ & is\_agn\_ctn                    & Is Compton-thin (CTN) AGN                                                 & \ldots       \\
$ 91$ & is\_agn\_ctk                    & Is Compton-thick (CTK) AGN                                                & \ldots       \\
$ 92$ & is\_agn                         & Is CTN or CTK AGN                                                         & \ldots       \\
$ 93$ & is\_agn\_focc                   & Is CTN or CTK AGN and hosts a BH according to the occupation fraction     & \ldots       \\
$ 94$ & log\_LX\_2\_10                  & $\logten$ of $2$--$10\,\mathrm{keV}$ intrinsic X-ray luminosity           & \unitlx      \\
$ 95$ & log\_FX\_2\_10                  & $\logten$ of $2$--$10\,\mathrm{keV}$ intrinsic X-ray flux at redshift     & \unitfx      \\
$ 96$ & log\_FX\_2\_7                   & $\logten$ of $2$--$7\,\mathrm{keV}$ intrinsic X-ray flux at redshift      & \unitfx      \\
$ 97$ & is\_optical\_type2              & Is optical \typeii \ac{AGN}                                               & \ldots       \\
$ 98$ & E\_BV                           & AGN $B-V$ color extinction                                                & mag          \\
$ 99$ & log\_L\_2\_keV                  & Monochromatic intrinsic $2\,\mathrm{keV}$ luminosity                      & \unitlnu     \\
$100$ & log\_L\_2500                    & Monochromatic intrinsic $2500\,\mathrm{keV}$ luminosity                   & \unitlnu     \\
$101$ & MBH                             & $\logten$ of black hole mass                                              & $M_\odot$    \\
$102$--$107$ & band\_tau                       & AGN damped random walk timescale (one per $ugrizy$)                       & days         \\
$108$--$113$ & band\_sf\_inf                   & AGN damped random walk structure function at infinity (one per $ugrizy$)  & ABmag        \\
$114$--$137$ & band\_point              & AGN or stellar band flux in the observer frame                            & $\mu$Jy      \\
$138$--$161$ & magabs\_band\_point      & AGN band absolute magnitude in the source's rest frame                    & ABmag        \\
$162$--$185$ & band\_total              & Total band flux in the observer frame                                     & $\mu$Jy      \\
$186$--$209$ & magabs\_band\_total      & Total band absolute magnitude in the source's rest frame                  & ABmag        \\
$210$ & pmracosd                        & Proper motion in $\mathrm{RA}\times \cos{\mathrm{Dec}}$                   & $\mathrm{mas}\,\mathrm{yr}^{-1}$ \\
$211$ & pmdec                           & Proper motion in Dec                                                      & $\mathrm{mas}\,\mathrm{yr}^{-1}$ \\
$212$ & is\_lsst\_any\_30s              & Flux is above LSST $30\,\mathrm{s}$   limit in any of the six bands       & \ldots       \\
$213$ & is\_lsst\_any\_10yr             & Flux is above LSST $10\,\mathrm{yr}$ limit in any of the six bands       & \ldots       \\
$214$ & is\_lsst\_all\_30s              & Flux is above LSST $30\,\mathrm{s}$   limit in all of the six bands       & \ldots       \\
$215$ & is\_lsst\_all\_10yr             & Flux is above LSST $10\,\mathrm{yr}$ limit in all of the six bands       & \ldots       \\
\hline
\end{tabular}
\tablefoot{%
  For each flux and absolute magnitude, the values are provided in the
  following bands: \ac{LSST}-$ugrizy$, Johnson $B$, Euclid $\YE\JE\HE$,
  Spitzer $\mathrm{IRAC}1$--$\mathrm{IRAC}4$, WISE $W1$--$W4$, and VISTA
  $YJHK_s$. We further provide two narrow-band fluxes in the \ac{UV}
  $1450\,\text{\AA}$ and \acl{MIR} $15\,\micron$.
}
\end{table*}

\section{Photometric catalog flux accuracy measurements}

We show the flux measurement accuracy from the \ac{AGILE} DR1 object table for
different classes in \cref{tab:flux_sigma_eta} (see
\cref{sec:photometric_accuracy}).

\begin{table*}[htbp]
  \caption{%
    Flux measurement accuracy across classes of objects and flux definitions in
    the \ac{AGILE} DR1 object catalog.
  }
  \label{tab:flux_sigma_eta}
\begin{tabular}{llrrrrrrrrr}
  \hline\hline
  & & \multicolumn{3}{c}{$20 \leq r < 22$} & \multicolumn{3}{c}{$22 \leq r < 24$} & \multicolumn{3}{c}{$24 \leq r < 26$} \\
Sample & Flux & $N$ & $\sigma_\mathrm{NMAD}$ & $\eta$ & $N$ & $\sigma_\mathrm{NMAD}$ & $\eta$ & $N$ & $\sigma_\mathrm{NMAD}$ & $\eta$ \\
\hline
AGN1 (<90\%) &    \texttt{psfFlux} &       $65$ &   0.208 &   0.477 &      $208$ &   0.169 &   0.341 &      $160$ &   0.159 &   0.381 \\
             &  \texttt{calibFlux} &       $65$ &   0.083 &   0.077 &      $208$ &   0.086 &   0.135 &      $160$ &   0.166 &   0.388 \\
             & \texttt{cModelFlux} &       $65$ &   0.084 &   0.092 &      $208$ &   0.068 &   0.135 &      $160$ &   0.134 &   0.306 \\
\hline
AGN1 (>90\%) &    \texttt{psfFlux} &       $77$ &   0.032 &   0.026 &       $64$ &   0.073 &   0.062 &       $13$ &   0.171 &   0.385 \\
             &  \texttt{calibFlux} &       $77$ &   0.053 &   0.065 &       $64$ &   0.059 &   0.078 &       $13$ &   0.164 &   0.308 \\
             & \texttt{cModelFlux} &       $77$ &   0.042 &   0.039 &       $64$ &   0.068 &   0.062 &       $13$ &   0.177 &   0.385 \\
\hline
AGN2 &    \texttt{psfFlux} &      $225$ &   0.476 &   0.871 &   $1635$ &   0.310 &   0.686 &   $4175$ &   0.221 &   0.499 \\
     &  \texttt{calibFlux} &      $225$ &   0.088 &   0.187 &   $1635$ &   0.072 &   0.141 &   $4178$ &   0.166 &   0.390 \\
     & \texttt{cModelFlux} &      $225$ &   0.063 &   0.147 &   $1635$ &   0.064 &   0.168 &   $4178$ &   0.135 &   0.315 \\
\hline
Galaxy &    \texttt{psfFlux} &   $7259$ &   0.555 &   0.956 &  $41\,327$ &   0.332 &   0.725 & $111\,608$ &   0.205 &   0.459 \\
       &  \texttt{calibFlux} &   $7259$ &   0.108 &   0.225 &  $41\,327$ &   0.069 &   0.138 & $111\,673$ &   0.165 &   0.393 \\
       & \texttt{cModelFlux} &   $7259$ &   0.070 &   0.159 &  $41\,326$ &   0.063 &   0.171 & $111\,669$ &   0.132 &   0.315 \\
\hline
Star &    \texttt{psfFlux} &   $2857$ &   0.003 &   0.007 &   $6228$ &   0.013 &   0.023 &   $8856$ &   0.061 &   0.112 \\
     &  \texttt{calibFlux} &   $2857$ &   0.008 &   0.028 &   $6228$ &   0.038 &   0.097 &   $8858$ &   0.158 &   0.384 \\
     & \texttt{cModelFlux} &   $2857$ &   0.004 &   0.015 &   $6228$ &   0.015 &   0.049 &   $8857$ &   0.070 &   0.163 \\
\hline
Pointlike &    \texttt{psfFlux} &   $2785$ &   0.003 &   0.001 &   $5071$ &   0.010 &   0.008 &  $17\,736$ &   0.109 &   0.225 \\
          &  \texttt{calibFlux} &   $2785$ &   0.007 &   0.001 &   $5071$ &   0.029 &   0.022 &  $17\,744$ &   0.185 &   0.431 \\
          & \texttt{cModelFlux} &   $2785$ &   0.004 &   0.001 &   $5070$ &   0.012 &   0.007 &  $17\,744$ &   0.102 &   0.213 \\
\hline
Extended &    \texttt{psfFlux} &   $7698$ &   0.546 &   0.934 &  $44\,391$ &   0.324 &   0.704 & $107\,076$ &   0.210 &   0.470 \\
         &  \texttt{calibFlux} &   $7698$ &   0.108 &   0.229 &  $44\,391$ &   0.071 &   0.146 & $107\,138$ &   0.161 &   0.385 \\
         & \texttt{cModelFlux} &   $7698$ &   0.071 &   0.161 &  $44\,391$ &   0.063 &   0.172 & $107\,133$ &   0.131 &   0.319 \\
\hline
\end{tabular}
\tablefoot{%
  The statistics are based on the $r$-band flux $F_r$. The normalized median
  absolute deviation is defined as $\sigma_\mathrm{NMAD} \equiv 1.48 \times
  \left|\, F_r - F_{r,\mathrm{truth}}\,\right| \,/\, F_{r,\mathrm{truth}}$, and
  the catastrophic outlier fraction $\eta$ is defined as the fraction of
  objects with $\left|\, F_r - F_{r,\mathrm{truth}}\,\right| \,/\,
  F_{r,\mathrm{truth}} > 0.15$, respectively. The magnitude limits refer to the
  truth catalog values.
}
\end{table*}

\end{appendix}

\end{document}